\documentstyle[aps,prc,preprint,epsf,tighten]{revtex}
\begin{document}
\baselineskip 16pt plus 2pt minus 2pt
\newcommand{\beq}{\begin{equation}}
\newcommand{\eeq}{\end{equation}}
\newcommand{\beqa}{\begin{eqnarray}}
\newcommand{\eeqa}{\end{eqnarray}}
\newcommand{\dfrac}{\displaystyle \frac}
\renewcommand{\thefootnote}{\#\arabic{footnote}}
\newcommand{\ve}{\varepsilon}
\newcommand{\krig}[1]{\stackrel{\circ}{#1}}
\newcommand{\barr}[1]{\not\mathrel #1}

\begin{titlepage}


\hfill TK 95 15

\hfill MKPH-T-95-07


\vspace{2.0cm}

\begin{center}

{\large  \bf {DISPERSION--THEORETICAL ANALYSIS OF THE NUCLEON
    ELECTROMAGNETIC FORMFACTORS}}\footnote{Work supported in part by the
    Deutsche Forschungsgemeinschaft (SFB 201).}

\vspace{1.2cm}

{\large P. Mergell$^{\ddag,\ast}$, Ulf-G. Mei\ss ner$^{\dag,\star}$,
D. Drechsel$^{\ddag,\diamond}$}

\vspace{1.0cm}

$^{\ddag}$Universit\"at Mainz, Institut f\"ur Kernphysik, J.-J.-Becher
Weg 45\\ D--55099 Mainz, Germany

\vspace{0.4cm}
$^{\dag}$Universit\"at Bonn, Institut f\"ur Theoretische Kernphysik, Nussallee
14-16,\\ D--53115 Bonn, Germany

\vspace{0.5cm}

$^\ast$electronic address: mergell@vkpmzp.kph.uni-mainz.de \\
$^\star$electronic address: meissner@pythia.itkp.uni-bonn.de \\
$^\diamond$electronic address: drechsel@vkpmzp.kph.uni-mainz.de

\end{center}

\vspace{0.7cm}

%
%
%
%
%
\begin{abstract}
\noindent Dispersion relations allow for a coherent description of the nucleon
electromagnetic form factors measured over a large range of momentum
transfer, $Q^2 \simeq 0 \ldots 35$ GeV$^2$.
 Including constraints from unitarity and perturbative QCD,
we present a novel parametrisation of the absorptive parts of the
various isoscalar and isovector nucleon form factors. Using the
current world data, we obtain results for the electromagnetic form
factors, nucleon radii and meson couplings. We stress the importance
of measurements at large momentum transfer to test the predictions of
perturbative QCD.
\end{abstract}


\vspace{2cm}

\vfill


\end{titlepage}

\section{Introduction and summary}
\label{sec:intro}
The electromagnetic structure of the nucleon as revealed in elastic
electron--nucleon scattering is parametrized in terms of the four
form factors $F_{1,2}^{p,n} (Q^2)$ (with $Q^2$ the squared momentum
transfer). The understanding of these form factors is of utmost
importance in any theory or model of the strong interactions.
Abundant data on these form factors over a large range of momentum
transfer already exist, and this data base will considerably improve in
the few GeV region when experiments at CEBAF will be completed and
analysed. In addition, experiments involving polarized beams and/or
targets are also performed at lower energies to give better data in
particular for the electric form factor of the neutron, but also the
magnetic proton and neutron ones. Such kind of experiments are under
way at MAMI, ELSA, MIT-Bates and other places. Clearly, theory has to
provide
a tool to interpret these data in a model--independent fashion.
Many years ago, dispersion theory was developed to extract these form
factors from elastic $ep$ and $ed$ scattering data (and others) (for some early
references, see e.g. \cite{chka} \cite{fego} \cite {drza}.). Such a
dispersion theoretical analysis is largely model--independent with the
exception of the absorptive parts of the form factors, which are often
parameterized in terms of  a few meson poles. A large class of the
early models of the
spectral functions related to the imaginary parts of $F_{1,2}^{p,n}
(Q^2)$ was based on the successful vector meson dominance (VMD)
hypothesis (for reviews, see \cite{zov} \cite{saku}) which states
that a photon couples to hadrons only
via  intermediate vector mesons. However, as already pointed out in
1959 by Frazer and Fulco \cite{frfu1} \cite{frfu2}, such an approach
is at odds with  general constraints from unitarity.
The two--pion continuum
has a pronounced effect on the isovector spectral functions on the
left wing of the $\rho$--resonance. This becomes
particularly visible in the determinations of the corresponding nucleon
radii. This effect  was  quantified by H\"ohler and Pietarinen
\cite{hopi} but has been neglected in most of the recent discussions of
the nucleons form factors, like in \cite{kk} \cite{gakr1} \cite{gakr2}
 \cite{du1} \cite{du2}. For an accurate determination of the isovector
 nucleon radii, the inclusion of the two--pion cut contribution is mandatory.
Another constraint comes from the  accurate
measurement of the neutron charge radius in very low-energy
neutron--atom  scattering  (for a recent update, see e.g.
 Leeb and Teichmeister  \cite{leeb}).
This leads to an additional normalization condition thus reducing the
number of free
parameters.\footnote{One can also consider the determination of the
electric charge radius of the proton by  Simon et al. \cite{simon} as
a further low--energy constraint. It is, however,
less  stringent as the one related to the neutron charge radius
 and will be discussed separately in section \ref{sec:prot}.}
Furthermore, in the framework of
perturbative QCD (pQCD), the behaviour of the pertinent form factors for
large momentum transfer can be inferred up to a model--dependent
normalisation \cite{brodsky}. This has lead to models which try to
synthezise aspects of low--energy hadron physics with the ones from pQCD,
see e.g. \cite{gakr1} \cite{gakr2} \cite{fuwa1} \cite{fuwa2}
\cite{kroll1} \cite{kroll2}. No clear
picture has yet emerged at which momentum transfer the asymptotic
behaviour really sets in. We stress here that any serious attempt to
describe the electromagnetic  structure of the nucleon
has to account for {\it all} these constraints.

The work presented here is an update and extension of the classical paper by
H\"ohler and collaborators \cite{hoeh76}. As in that paper, we
determine the $\rho$--meson contribution to the isovector  form
factors from extended unitarity as discussed by Frazer and Fulco
\cite{frfu2}. While we use the old $\pi N$ partial waves extrapolated
to the time--like regime \cite{hopipi}, we update the pion form factor
to account for the measured $\rho-\omega$ mixing \cite{barkow}.
In addition, since we are dealing with what is called an
ill-posed problem, we restrict the number of additional pole terms in
the isovector and isoscalar channel to a minimum (see the discussion
below). We also implement the
large--$Q^2$ behaviour dictated by pQCD, similar in spirit to the
work of Furuichi and Watanabe \cite{fuwa1} \cite{fuwa2}. However, as we
demonstrate later on, their choice of the spectral functions is not
compatible with pQCD. What one ends up with is a set of
superconvergence relations which reduce the number of free parameters.
These superconvergence relations can be implemented in a variety of
ways. Therefore, we will discuss two versions of the spectral
functions  with the proper asymptotic behaviour guided by simplicity
and the notion of separating the hadronic from the pQCD contribution.
A fit to the accessible data then allows not only
to extract the nucleon electromagnetic form factors, but also the
nucleon radii and nucleon--meson coupling constants. In particular, we
also find a large coupling of the $\Phi$--meson to the nucleon,
seemingly in contradiction with the OZI--rule as stressed in the
work by Genz and
H\"ohler \cite{geho} and, more recently, by Jaffe \cite{bob}.
It should be already noted here that this large $\Phi$--nucleon
coupling can be understood in terms of coupling of the photon to the kaon cloud
surrounding the nucleon \cite{gakr3}. Our approach
furthermore sheds light on the possible onset of pQCD and the regime
where data are urgently called for. For a status review on the
determination of the nucleon electromagnetic form factors, we refer to
H\"ohler \cite{ho93}, Milner \cite{richard} and Bosted \cite{bostedr}.

\medskip

\noindent The pertinent results of this investigation can be summarized as
follows:

\begin{enumerate}

\item[(i)] Including all the constraints discussed and using the
current world data, we have found a new fit to the nucleon
electromagnetic form factors. Besides the two--pion continuum and the
pQCD contributions, we have three isoscalar and three isovector poles.
This is the minimum number required by the data. We also stress that
it is mandatory to {\it simultaneously} fit the proton and the neutron data.

\item[(ii)] The form factors $F_1^{(s)} (t)$ and $F_2^{(v)} (t)$ show
a marked dipole structure. The reasons is that the two lowest poles in
the respective channels are not too far separated in mass and appear
with residua of the same magnitude but different in sign. In the isoscalar
case, this is due to the
closeness of the $\omega$ and $\Phi$ mesons \cite{hoeh76}. The novel
feature of our study is that the first pole above the two--pion
continuum can be identified with the $\rho ' (1450)$ leading to the
dipole structure in $F_2^{(v)} (t)$. We furthermore find the
third isoscalar and the second isovector pole at $M_{S'} = 1.60$~GeV
and $M_{\rho ''} = 1.65$~GeV. Only the third isovector pole cannot be
identified with a physical state.

\item[(iii)] We have given a new determination of the electric and
magnetic radii of the proton and the neutron. We find $r_E^p =
0.847$~fm, $r_M^p = 0.853$~fm and $r_M^n = 0.889$~fm, all with an
uncertainty of about 1\%.
These are consistent with the findings of ref.\cite{hoeh76} with the
exception of the neutron magnetic radius. The difference is mainly due
to the neglect of the superconvergence relation related to the
asymptotic behaviour of $F_2^{(v)}$ in \cite{hoeh76}.

\item[(iv)] We have shown that the accurate determination of the
proton charge radius in ref.\cite{simon} is only consistent
with the dispersive analysis  within one standard deviation.
Using the central value of ref.\cite{simon} for $r_E^p$, one can{\it
  not}
{\it simultaneously} fit the proton and the neutron data.

\item[(v)] We have determined the vector ($g_1$)
and tensor ($g_2$) couplings of the
$\omega$ and the $\Phi$ mesons. The $g_1$ are slightly larger than in
Ref.\cite{hoeh76} while the $g_2$ are of comparable size. We stress
again that the large $\Phi NN$ coupling does not indicate a violation
of the OZI--rule but rather  accounts for the neglect of the $\bar
K K$ continuum.

\item[(vi)] In our best fit, the parameter $\Lambda^2$, which is a
measure of the boundary between the hadronic (pole) and quark (pQCD)
contributions, comes out to be $\Lambda^2 \simeq 10$~GeV$^2$. Only in
the case of the proton magnetic form factor one has data at suffciently
large momentum transfer to possibly  test the predictions of pQCD. Our
conclusion is that even in that case more data are needed to really
see the asymptotic behaviour $Q^4 G_M^p (Q^2) \to$~constant
(modulo logarithms). For all
form factors there is still a sizeable hadronic contribution in the
momentum transfer range between $Q^2 =10 \ldots 20$~GeV$^2$.
In particular, the data
available for the ratio $Q^2 F_2 (Q^2) / F_1 (Q^2) $ are at too
low momentum transfer to indicate any scaling.

\item[(vii)] The fits are completely insensitive to the number of
flavors entering via the anomalous dimension appearing in
the asymptotic behaviour of the
form factors and to the $\rho-\omega$
mixing which comes in via the calculation of the isovector two--pion
unitarity correction.

\item[(viii)] More accurate data at {\it low}, {\it intermediate} and
(very) {\it large} momentum transfer are called for to further tighten
the constraints on the radii and coupling constants and to really test
the pQCD predictions.

\end{enumerate}

The manuscript is organized as follows. In section \ref{sec:con}
we discuss the basic concepts underlying the dispersion--theoretical
analysis of the nucleon electromagnetic form factors. In particular,
we summarize the various constraints which have to be implemented to
obtain a consistent picture. These are the inclusion of the two--pion
continuum in the isovector spectral functions (sect. \ref{sec:exuni}), the
determination of the neutron charge radius from very low energy
neutron--atom scattering (sect. \ref{sec:neut}) and the behaviour of the
various form factors as given by pQCD (sect. \ref{sec:pqcd}). We also discuss
another less stringent constraint related to the proton charge radius
(sect. \ref{sec:prot}). Section \ref{sec:fitf} contains the form factor
parametrizations we will use. We present two forms, one in which the
pQCD constraints are implemented in the most simple fashion and
another one, which allows for a separation between the
hadronic (pole) and the pQCD terms already on the level of the
spectral functions. The results are presented and discussed in section
\ref{sec:res}. The appendices~A and B contain some technicalities, and in
appendix~C, we give the explicit parametrization of our best fit.

\section{Basic concepts}
\label{sec:con}
In this section, we assemble all tools necessary for the construction
of the fit functions for the spectral distributions discussed in
section \ref{sec:fitf}. We also supply the basic definitions and notations
underlying our analysis. In particular, we discuss the various low and
high energy constraints which have to be imposed and allow
to reduce the number of free
parameters. This material is mostly not new but necessary to keep the
manuscript self--contained. The reader familiar with it is invited to
skip this section.

\subsection{Definition of the nucleon electromagnetic
 form factors}
\label{sec:def}
The matrix--element of the electromagnetic (em) current ${\cal J}_\mu$
sandwiched between nucleon states can be expressed in terms
of two form factors,
\beq
<N(p')\,| \, {\cal J}_\mu \, | \, N(p)> =
e \, \bar{u}(p') \biggl\lbrace \, \gamma_\mu \,
F_1 (t) + \dfrac{i}{2m_N} \, \sigma_{\mu \nu} \, Q^\nu \, \, F_2 (t)
\biggr\rbrace \, u(p) \, \, \,  ,
\eeq
with $m_N$ the nucleon mass, $t = (p'-p)^2 = -Q^2$ the invariant
momentum transfer squared and $t<0$ in the space--like region.
$F_1 (t)$ and $F_2(t)$ are called the Dirac and the Pauli form factor,
respectively. They are normalised at $t=0$ as
\beq
F_1^p(0) = 1 \, , \quad F_1^n(0) = 0 \, , \quad F_2^p(0) =  \kappa_p \, ,
\quad F_2^n(0) = \kappa_n \, \, , \eeq
with $\kappa_p$ ($\kappa_n$) the anomalous magnetic moment of the
proton (neutron) in units of the nuclear magneton, $\mu_N = e/2m_N$.
For later use, we also need the isospin decomposition, i.e. the
response of the nucleon to the isoscalar (denoted by the superscript
'$(s)$') and isovector (denoted by the superscript '$(v)$') components of
the electromagnetic current,
\beq
F_i^{(s)} = \dfrac{1}{2} (F_i^p + F_i^n) \, , \quad
F_i^{(v)} = \dfrac{1}{2} (F_i^p - F_i^n) \, , \quad
(i = 1,2) \, , \eeq
with the normalisation
\beq F_i^I (0) = \nu_i^I \, \, , \quad i=1,2 \, ; \quad I = (v),(s) \,  \, ,
\eeq
where
\beq
\nu_1^{(s)} = \nu_1^{(v)} =\dfrac{1}{2} \, \, , \, \,
\nu_2^{(s)} = \frac{1}{2}(\kappa_p + \kappa_n ) \, \, , \, \,
\nu_2^{(v)} = \frac{1}{2}(\kappa_p - \kappa_n ) \, \, .
\label{nu} \eeq
One can also introduce the so--called Sachs form factors
$G_{E,M}^{p,n} (Q^2)$ related to the $F_1(Q^2) $ and $F_2 (Q^2)$ via
\beq
G_E(Q^2) = F_1(Q^2) - \tau F_2(Q^2) \, , \quad
G_M(Q^2) = F_1(Q^2) + F_2(Q^2) \, , \quad \tau = \dfrac{Q^2}{4m_N} \,
. \eeq
In the Breit frame, $G_E$ and $G_M$ are nothing but the
Fourier--transforms of the charge and the magnetization distribution,
respectively.

The slope of the form factors at $t=0$ is conventionally expressed in
terms of a nucleon radius $<r^2>^{1/2}$,
\beq
F(t) = F(0) \, \biggl( 1 + \dfrac{1}{6}<r^2> \, t + \ldots \biggr) \,
\eeq
which is rooted in the non--relativistic description of the scattering
process, in which a point-like charged particle interacts with a
given charge distribution $\rho(r)$. The mean square radius of this
charge distribution is given by
\beq
<r^2> = \int_0^\infty dr \, 4 \pi r^2 \rho(r) = -
\dfrac{6}{F(0)} \dfrac{dF(Q^2)}{dQ^2} \biggr|_{Q^2 = 0} \, \, .
\label{defrad}  \eeq
Eq.(\ref{defrad}) can be used for all form factors except $G_E^n$ and
$F_1^n$ which vanish at $t=0$. In these cases, one simply drops the
normalization factor $1/F(0)$ and defines e.g. the neutron charge radius
via
\beq
<r^2>_n^{\rm ch}  \, \, = -6 \, \dfrac{dG_E^n(Q^2)}{dQ^2}
\biggr|_{Q^2 = 0} \, \, . \eeq
To conclude this section, we remark that the slopes of $G_E^n$ and
$F_1^n$ are related via
\beq
\dfrac{dG_E^n(Q^2)}{dQ^2} \biggr|_{Q^2 = 0} = \dfrac{dF_1^n(Q^2)}{dQ^2}
\biggr|_{Q^2 = 0} - \dfrac{F_2^n (0)}{4m_N^2} \, \, .
\label{foldy}   \eeq
The second term in eq.(\ref{foldy}) is called the Foldy term. It gives
the dominant contribution to the slope of $G_E^n$.

\subsection{Dispersion relations for the nucleon em form factors}
\label{sec:disp}
Let $F(t)$ be a generic symbol for any one of the four nucleon em form
factors $F_{1,2}^{(v,s)}(t)$. We assume the validity of an unsubtracted
dispersion relation of the form\footnote{Since the normalization $F(0)$
  is known, one could also work with once--subtracted dispersion relation.}
\beq
F(t) = \dfrac{1}{\pi} \, \int_{t_0}^\infty \dfrac{{\rm Im} \,
  F(t')}{t'-t} \, dt' \, \, \, .
\label{disp} \eeq
The spectral function Im~$F(t)$ is different from zero along the
cut from $t_0$ to $\infty$ with $t_0 = 4 \, (9) \, M_\pi^2$ for the
isoscalar (isovector) case and $M_\pi$ denotes the pion mass. The
proof of the validity of such dispersion relations in QCD has not yet
been given \cite{oehme}. Eq.(\ref{disp}) means that the em structure
of the nucleon is entirely determined from its absorptive behaviour.
Data for $F(t)$ are given for $t<0$. If these
would be infinitively precise, the continuation
to other values of $t> 4M_\pi^2$
would be unique.   However, the available data have certain error bars
which make the continuation unstable as detailed in the review
\cite{cpss}. This is what one calls an ill--posed problem.
Therefore, one needs some additional assumptions which should have a
physics motivation. We follow here the
prescription due to Sabba--Stefanescu \cite{sab} and require that the
spectral functions should not have more than the minimum number of oscillations
and parameters required by the data. To make this more specific, let
us consider as an example the rather successful dipole fit,
\beq
G_E^P (t) \simeq \mu_p G_M^p  (t) \simeq\mu_n G_M^n  (t) \simeq
G_D (t) \, , \quad G_D(t) = (1 - t/0.71 \,{\rm GeV}^2)^{-2} \, \, ,
 \label{dipole}
\eeq
where $\mu_p$ and $\mu_n$ are the magnetic moments of the proton and
the neutron, respectively.
Such a form is only possible if the corresponding spectral functions
contain at least two pole terms with opposite signs.

Of particular interest is the VMD approach in which the spectral
functions are approximated by a few vector meson poles, namely the $\rho,
\ldots$ in the isovector and the $\omega, \Phi, \ldots$ in the
isoscalar channel, in order. In that case, the form factors take the
form
\beq
F_i^{(I)} (t) = \sum_V \dfrac{a_i^{(I)}}{M^2_{(I)}-t} \, \quad i = 1,2
\, ;
\,  \quad I = v,s  \, \, , \label{vmd} \eeq
with
\beq
a_i^V = \dfrac{M^2_V}{f_V} \, g^{VNN}_i \, , \quad V = \rho, \omega,
\Phi, \ldots \, \, , \, i =1,2 \, \, .
 \label{vmdc}  \eeq
In many fits, the mass parameters $M_V$ are taken from the known meson
masses and the strength parameters $a_i^V$ are fitted. This in turn
leads to a determination of the various vector--meson--nucleon coupling
constants, $g^{VNN}_i$. Note, however,
that such a procedure becomes increasingly
arbitrary for higher mass excitations.
One defines the ratio of the tensor to vector coupling, $\kappa_V$, via
\beq
\kappa_V =  \dfrac{g^{VNN}_2}{g^{VNN}_1} \, \, \, .
\label{kappa}   \eeq
One expects e.g. $\kappa_\rho$ to be large ($\sim 6$) and
$\kappa_\omega$ to be small ($\sim 0$). In strict VMD, one has
$\kappa_\rho = \kappa_p - \kappa_n = 3.71$ and
$\kappa_\omega = \kappa_p + \kappa_n = -0.12$.
The coupling constants $f_V$ are known from the widths of the leptonic decays
$V \to e^+ e^-$, i.e.
\beq
\dfrac{f_V^2}{4 \pi} = \dfrac{\alpha^2}{3} \dfrac{M_V}{\Gamma (V \to
    e^+ e^-)} \, \, \, . \eeq
Clearly, such pole terms contribute to the spectral functions in terms
of $\delta$--functions,
\beq
{\rm Im}F^V_i (t) = \pi\,  a_i^V \, \delta(t - M_V^2) \, \, . \eeq
Of course, there are other contributions related to intermediate
states like $n \pi$ ($n \ge 2)$, $N \bar{N}$, $K \bar{K}$ and so on.
As we will discuss in section \ref{sec:exuni}, of these the $2\pi$
intermediate states play the most important role.

\subsection{Constraints from unitarity}
\label{sec:exuni}
Here, we briefly summarize what is known about the contribution of the
two--pion continuum to the isovector spectral functions and how this
should be implemented. The unitarity relation of Frazer and Fulco
\cite{frfu2} determines the isovector spectral functions from $t=
4M_\pi^2$ to $t \simeq 50 M_\pi^2 \simeq 1$ GeV$^2$ in terms of the
pion form factor $F_\pi (t)$ and the P--wave $\pi \pi N \bar{N}$
partial wave amplitudes, cf. Fig.1. We use here the form
\beq
{\rm Im}F_i^{(v), 2\pi} (t) = \dfrac{q_t^3}{\sqrt{t}} \, |F_\pi (t)|^2
  \, J_i (t) \, , \quad i=1,2
\label{im2pi} \eeq
with $q_t = \sqrt{t/4 - M_\pi^2}$ and the functions $J_i (t)$ are related to
the P--wave $\pi N$ partial waves in the $t$--channel, $f_{\pm}^1
(t)$, via
\beq
f_{\pm}^1 (t) = F_\pi (t) \, J_\pm  (t)  \, \, ,
\eeq
in the conventional isospin decomposition. The $J_i (t) $ are
tabulated in \cite{bible}. For the pion form factor, we use the
recent parametrization of Barkow et al. \cite{barkow} which takes into
account $\rho-\omega$ mixing,
\beq
F_\pi (t) = F_\pi^\rho (t) \dfrac{1+ \alpha_\omega F_\pi^\omega (t)}{1
  + \alpha_\omega} \, \,
\label{fpi}  \eeq
with $\alpha_\omega$ the mixing parameter, $\alpha_\omega = 0.0038$
\cite{barkow}. The functions $F_\pi^V (t) $ $(V = \rho , \omega)$ are
of the
standard Gounaris--Sakurai form \cite{gousa}. A typical result for the
corresponding isovector spectral functions (weighted with $1/t^2$) is
shown in Fig.2. One notices the strong enhancement close to two--pion
threshold. The reason for this behaviour has been known for long. The
partial waves $f_\pm^1 (t)$ have a branch point singularity on the
second sheet (from the projection of the nucleon pole terms) located
at
\beq
t_c = 4 M_\pi^2 -M_\pi^4 / m_N^2 = 3.98 \, M_\pi^2 \, \, ,
\eeq
very close to the physical threshold at $t_0 = 4 M_\pi^2$. The
isovector form factors inherit this singularity and the closeness to
the physical threshold leads to the pronounced enhancement. Note that
in the VMD approach this spectral function is given by a
$\delta$--function peak at $M_\rho^2 \simeq 30 M_\pi^2$ and thus the
isovector radii are strongly underestimated if one neglects the
unitarity correction \cite{hopi} as can be seen from the formula
\beq
<r^2>_i^{(v),2\pi} = \dfrac{6}{\pi}
 \int_{4M_\pi^2}^{50 M_\pi^2} \dfrac{dt'}{{t'}^2} \, {\rm Im}
  F_i^{(v), 2\pi} (t') \, \, . \eeq
Consequently, in the isovector channels one should not use a simple
$\rho$ pole but rather work with the spectral functions as given by
eq.(\ref{im2pi}). This is the procedure adopted here. To speed up the
numerical calculations, one can fit the corresponding two--pion
contributions to the isovector form factors via
(as it was done in \cite{hoeh76})
\beq
F_i^{(v), 2\pi} (t) = F_i^\rho (t) = \dfrac{a_i^\rho + b_i^\rho \bigl(
    1 - t/ c_i^\rho \bigr)^{-2/i}}{2 \, ( 1 - t/d_i^\rho)} \,  , \quad i=1,2
\label{fit2pi} \eeq
with $
a_1^\rho = 1.0317$, $a_2^\rho = 5.7824$, $b_1^\rho = 0.0875$,
$b_2^\rho = 0.3907$, $c_1^\rho = 0.3176$, $c_2^\rho = 0.1422$,
$d_1^\rho = 0.5496$ and $d_2^\rho = 0.5362$. We have also
performed fits with the exact representation eq.(\ref{im2pi}) which
lead to the same results as the use of the form eq.(\ref{fit2pi}).
To end this section, we remark that the form eq.(\ref{im2pi}) is exact
below four--pion threshold, $t = 16 M_\pi^2$ and we stop at $t_{\rm
  up} = 50 M_\pi^2$ since the pion form factor shows some structure on
the right wing of the $\rho$--peak which can not simply be fitted by a
superposition of Gounaris--Sakurai functions. In the isoscalar
channel, it is believed that
the pertinent spectral functions rise smoothly
from the three--pion threshold  to the $\omega$ peak,
i.e. that there is no pronounced effect from the three--pion cut on
the left wing of the $\omega$--resonance (which also has a much
smaller width than the $\rho$).
Chiral perturbation theory \cite{gss} \cite{bkmrev}
or an investigation of the spectral functions
related to the process $N \bar{N} \to 3 \pi $
could be used to settle this issue. The one loop
calculation of the isovector nucleon form factors indeed shows the
unitarity correction on the left wing of the $\rho$ \cite{gss}. For
the isoscalar form factors, a two loop calculation of the pertinent
imaginary parts would reveal whether there is some enhancement around
$t = 9 M_\pi^2$ or justify the common assumption that one has a
smooth isoscalar spectral functions driven by the $\omega$ at low $t$.

\subsection{Constraints from low--energy neutron--atom scattering}
\label{sec:neut}
There exists a large number of experiments trying to determine the
neutron--electron scattering length $b_{ne}$ from low--energy neutron--atom
scattering. The interest in this quantity stems from its direct
relation to the mean square radius of the neutron,
\beq
b_{ne} =  \dfrac{\alpha m_n}{3} \, <r^2>_n^{\rm ch} \, \, , \eeq
or
\beq
\dfrac{dG_E^n (Q^2)}{dQ^2}\biggr|_{Q^2 =0} = - 14.39 \, [{\rm fm}] \, b_{ne}
\, \, , \label{neurad} \eeq
with $\alpha = e^2 / 4\pi = 1/ 137.036$ the fine structure constant.
For an early review, see e.g. Foldy \cite{llfol}.
There has been some controversy about the actual value of $b_{ne}$
over the years as discussed in \cite{leeb}. We use here the most
recent value, $b_{ne} = (-1.308 \pm 0.05) \cdot 10^{-3}$ fm
\cite{kopek} which leads to (adding systematic and statistical errors
in quadrature)
\beq
-6\dfrac{dF_1^n (Q^2)}{dQ^2}\biggr|_{Q^2 =0} =
(0.0136 \pm 0.0043) \, {\rm fm}^2 \, \, , \label{neuc}  \eeq
using eq.(\ref{foldy}). Notice that due to the large cancellation
between the Foldy term and the slope of $G_E^n$, the slope of $F_1^n$
is very sensitive to the value of $b_{ne}$. If one uses e.g. the older
value of Koester et al. \cite{koest}, $b_{ne} = (-1.32 \pm 0.04)
\cdot 10^{-3}$ fm, one finds $-6 dF_1^n/dQ^2  = (0.0126 \pm
0.0035)$~fm$^2$. The slope of $F_1^n (Q^2)$
as given in eq.(\ref{neuc}) will be imposed on all our fits.

\subsection{Constraints from low--energy electron--proton scattering}
\label{sec:prot}
Simon et al. \cite{simon} have presented a precise measurement and
analysis of their and other existing data for elastic electron--proton
scattering in the range $Q^2 = 0.005 \ldots 0.055$ GeV$^2$. They have
performed a fit of the type
\beq
G_E^p (Q^2) = a_0 \biggl( 1 + Q^2 \dfrac{a_1}{a_0} + Q^4
\dfrac{a_2}{a_0} \biggr)  \, \, , \label{sifit}
\eeq
with adjustable parameters $a_0, a_1, a_2$. The
proton charge radius is thus given by
\beq
\dfrac{1}{G_E^p (0)} \dfrac{dG_E^p (Q^2)}{dQ^2}\biggr|_{Q^2 =0}
= \dfrac{a_1}{a_0} =
- \dfrac{<r^2_E>_p}{6 a_0} \, \, . \eeq
For the data in the low--energy region, the contribution of the $Q^4$
term to the proton electric form factor is marginal ($< 0.3 \%$). This
leads to an rather accurate value for $<r^2_E>_p$,
\beq
<r^2_E>_p = (0.862 \pm 0.012) \, \, {\rm fm}^2  \, \, .
\label{simon}
\eeq
With that constraint, the authors of \cite{simon} performed a four
pole fit (with two masses fixed at $M_\rho = 0.765$ GeV and $M_{\rho
  '} = 1.31$ GeV) to the available data for the proton
electric and magnetic form factors up to $Q^2 \simeq 5$ GeV$^2$. This
allowed to reconstruct the spectral function Im$G_E^v (t)$. It is in
agreement with the one of ref.\cite{hopi} with the exception of the
region around $t = 7 M_\pi^2$. One could now argue that the dispersive
analysis to be performed should be constrained by the value of the proton
radius, eq.(\ref{simon}). In that case one would, of course, have to
exclude the low--energy data ($Q^2 < 0.6$ GeV$^2$) from the data
basis.
So as not to exclude the
possibility of new and more precise data also for $G_E^p$ at low
$Q^2$, we will  perform a set
of fits with all data included and another set with the constraint
eq.(\ref{simon}) imposed and the corresponding data removed.
We consider this as a good measure of the
accuracy in determining the various nucleon radii from  a
dispersion--theoretical analysis.

\subsection{Constraints from perturbative QCD}
\label{sec:pqcd}
Perturbative QCD allows to constrain the behaviour of the nucleon em
form factors for large momentum transfer, $Q^2 = -t \to \infty$. In its
most simple fashion, the so--called quark counting rules \cite{brfa}
give the leading power in the large--$Q^2$ fall--off of the form
factors by counting the number of gluon exchanges which are necessary
to distribute the large photon momentum equally to all
constituents. In the limit $-t \to \infty$ this leads to
\beq
 (-t)^{i+1} \, F_i (t) \to {\rm constant} \, \, , \quad i = 1,2
\eeq
which in turn translates into a set of superconvergence relations for
the spectral functions,
\beq
\dfrac{1}{\pi} \int_{t_0}^\infty dt' \, {\rm Im}F_1 (t') = 0 \, \, ,
\eeq
\beq
\dfrac{1}{\pi} \int_{t_0}^\infty dt' \, {\rm Im}F_2 (t') =
\dfrac{1}{\pi} \int_{t_0}^\infty dt' \, t' \, {\rm Im}F_2 (t') = 0 \, \, ,
\eeq
for both the proton and the neutron. These arguments have been
sharpened in \cite{brodsky}. There, it was shown that the magnetic
Sachs form factor takes the form
\beq
G_M(Q^2) = \, C \, \dfrac{\alpha_s^2 (Q^2)}{Q^4} \biggl[ \ln \biggl(
\dfrac{Q^2}{Q_0^2}\biggr) \biggr]^{\frac{2}{3 \beta}} \, \, ,
\label{gmasy} \eeq
with $\alpha_s (Q^2)$ the running strong coupling constant,
\beq
\alpha_s (Q^2) = \dfrac{4 \pi}{\beta \, \ln(Q^2/Q_0^2)} \, , \quad
\beta = 11 - \dfrac{2}{3} N_f \, \, .
\eeq
Here, $N_f$ is the number of flavors, $\beta$ the QCD
$\beta$--function to one loop and $Q_0 \simeq \Lambda_{\rm QCD}$. The
constant $C$ in eq.(\ref{gmasy}) is model--dependent and subject of
much controversy. Its explicit form is not needed here. Taking
furthermore into account that any helicity--flip leads to an extra
$1/Q$ factor, one finds
\beq
F_i (t) \to (-t)^{-(i+1)} \, \biggl[ \ln\biggl(\dfrac{-t}{Q_0^2}\biggr)
\biggr]^{-\gamma} \, , \quad \gamma = 2 + \dfrac{4}{3\beta}
\, \, , \quad i = 1,2 \, \, .
\label{fasy}  \eeq
The anomalous dimension $\gamma$ depends weakly on the number of
flavors, $\gamma = 2.148$, $2.160$, $ 2.173$ for $N_f = 3$, $4$,
$5$, in order.
In what follows, we will construct spectral functions which lead
exactly to the large--$t$ behaviour as given in eq.(\ref{fasy}).

\section{Parametrization of the form factors}
\label{sec:fitf}
We have now assembled all tools to construct parametrizations of the
nucleon form factors subject to the constraints discussed in the
previous section. We will present two such parametrizations. The first
one, which we label 'multiplicative', is guided by simplicity. For the
second one, called 'additive', the spectral function is split in a way
which allows for a clearer separation between the hadronic (pole) and
the pQCD contributions.

\subsection{Multiplicative parametrization}
\label{sec:multfit}
Consider the functions
\beq
F_i^{(s)} (t) = \tilde{F}_i^{(s)} (t) L(t) = \biggl[ \sum_s
\dfrac{a_I^{(s)} \, L^{-1}(M^2_{(s)})}{M^2_{(s)} - t }\biggr] \, \biggl[ \ln
\biggl( \dfrac{\Lambda^2 - t}{Q_0^2} \biggr)\biggr]^{-\gamma} \, \, ,
\label{ffism} \eeq
\beq
F_i^{(v)} (t) = \tilde{F}_i^{(v)} (t) L(t) = \biggl[
\tilde{F}_i^\rho (t) + \sum_v
\dfrac{a_I^{(v)} \, L^{-1}(M^2_{(v)})}{M^2_{(v)} - t }\biggr] \, \biggl[ \ln
\biggl( \dfrac{\Lambda^2 - t}{Q_0^2} \biggr)\biggr]^{-\gamma} \, \, ,
\label{ffivm} \eeq
\beq
L(t) \equiv \biggl[ \ln
\biggl( \dfrac{\Lambda^2 - t}{Q_0^2} \biggr)\biggr]^{-\gamma} \, \, ,
\label{deflt}
\eeq
where the parameter $\Lambda$ can be considered as a measure of the
onset of the asymptotic behaviour. The value of  $Q_0$ is strongly
correlated to the one of $\Lambda$ in the actual fits, we choose its
value to be in the few hundred MeV region (as discussed in
section \ref{sec:res}), $Q_0 \sim \Lambda_{\rm QCD}$.
We take into account three poles in the isovector $(v)$
channels and denote these by $\rho ', \rho ''$ and $\rho
'''$.\footnote{This notation, however, should not imply a priori that these
poles have to be identified with higher mass excitations of the
$\rho$. This topic will be taken up again when we discuss the actual
results of our fits.}
The $\rho$ contribution is fixed by extended unitarity as discussed in
section \ref{sec:exuni},
\beq
\tilde{F}_i^\rho (t) = \dfrac{1}{\pi} \int_{4M_\pi^2}^{50M_\pi^2} dt'
\, \dfrac{{\rm Im}F_i^\rho (t')}{L(t') (t' - t)} =
\dfrac{ a_i^\rho L^{-1}(M_a^2) + b_i^\rho L^{-1}(M_b^2) \bigl( 1 - t /
  c^\rho_i \bigr)^{-2/i}}{2 \bigl( 1 - t /  d^\rho_i \bigr)} \, \, ,
\label{lt} \eeq
with the mass parameters $M_a$ and $M_b$ determined
from a fit to $\tilde{F}_i^\rho
(t)$ and the $a_i^\rho, b_i^\rho, c_i^\rho, d_i^\rho$ given after
eq.(\ref{fit2pi}). We find $M_a^2 = 0.5$~GeV$^2$ and
$M_b^2 = 0.4$~GeV$^2$. In the isoscalar channels, we have three meson
poles, the first two with fixed masses at $M_\omega = 0.782$ GeV and
$M_\Phi = 1.02$ GeV plus one heavier state denoted by $S'$.
The contribution of $L(t)$ for momenta smaller than $\Lambda$ is
balanced by the appropriate normalization factors at the meson poles,
$ L^{-1}(M^2_I)$, $I=(v),(s)$.

The analytic structure of the functions $F_i^I (t)$ in the complex
$t$-plane is shown in Figs.3a,b. In the isovector case (Fig.3b),
$I = (v)$, the
two--pion continuum leads to a cut from $t_0 = 4M_\pi^2$ to $t_{\rm
  up} = 50 M_\pi^2$. In addition, there are poles at $t = M_{\rho'},
M_{\rho''},M_{\rho'''}$ which to some extent  compensate the neglect of
cuts related to higher mass continua like $4 \pi$, $N \bar{N}$ and so
on (with positive G--parity). At larger values
of $t$ along the real axis, there is a
singularity at $t_{\rm sing} = \Lambda^2 -
Q_0^2$ from the logarithm plus a cut
from $t_{\rm sing}$ to $\Lambda^2$ since the exponent $\gamma$ appearing in
$L(t)$, eq.(\ref{deflt}), is rational. Finally, there is the right--hand cut
from $\Lambda^2$ to $\infty$. The only difference in the isoscalar
case (Fig.3a), $I = (s)$, is that one has
three poles at $M_\omega^2$, $M_\Phi^2$
and $M_{S'}^2$. The latter two simulate cuts related to
multi--pion and other intermediate states with negative G--parity.
The singularity at $t_{\rm sing}$ has no physical significance, it is
related to the way in which the large---$t$ behaviour is enforced. We will
come back to this point in the next subsection.

The normalization conditions of the form factors at $t=0$ take the
form
\beq
\rho_i^{(-1)} \, \delta_{I, (v)} + \sum_I \, a_i^I \,
\dfrac{L^{-1}(M^2_I)}{M^2_I} = \nu_i^I \, L^{-1} (0) \, \,
\label{normm} \eeq
with the $\nu_i^I$ given in eq.(\ref{nu}) and \cite{fuwa1}
\beq
\rho_i^{(k)} = \dfrac{1}{\pi} \int_{4M_\pi^2}^{50M_\pi^2} dt' \, {t'}^k
\, {\rm Im}F_i^\rho (t') L^{-1} (t') = \dfrac{1}{2} \biggl( a_i^\rho \,
L^{-1}(M_a^2) \, (d_i^\rho)^{k+1} + b_i^\rho \, L^{-1}(M_b^2) \,
\delta_{k,-1} \biggr) \, \, . \label{rho} \eeq
The conditions on the
neutron charge radius, eq.(\ref{neuc}), and on the proton charge radius,
eq.(\ref{simon}), translate into
$$ \dfrac{dF^I_1}{dt}(0) \, L(0) = \sum_I \dfrac{a_1^I
  L^{-1}(M_I^2)}{M_I^2} \biggl( \dfrac{1}{M_I^2} +
\dfrac{\gamma}{\Lambda^2 L^{-1/\gamma}(0)} \biggr) $$
\beq + \delta_{I,(v)}
\biggl[ \biggl( a_1^\rho L^{-1}(M_a^2) +b_1^\rho L^{-1}(M_b^2) \biggr)
\biggl(\dfrac{1}{d_1^\rho}
+ \dfrac{\gamma}{2 \Lambda^2 L^{-1/\gamma}(0)} \biggr) + 2
\dfrac{b_1^\rho L^{-1}(M_b^2)}{c_1^\rho} \biggr]
\, \, , \,\, \, I = (v),(s) \, .
\label{neuprom} \eeq
If one only imposes the value of the neutron charge radius,
one has to take the appropriate linear combination of
the two conditions given in eq.(\ref{neuprom}). We stress again that
this latter case should be considered as more realistic.
The implementation of the pQCD constraints,
eq.(\ref{fasy}), via the logarithmic function $L(t)$ is similar to the
approach taken in \cite{gakr1,gakr2,gakr3}. The following
superconvergence relations have to be fulfilled so that the proper
pQCD asymptotics emerges \cite{fuwa1}
\beqa
\rho_i^{(0)} \, \delta_{I, (v)} + \sum_I \, a_i^I \,L^{-1}(M^2_I)
=0 \, \, ,\nonumber \\
\rho_2^{(1)} \, \delta_{I, (v)} + \sum_I \, M^2_I \, a_2^I \,L^{-1}(M^2_I)
= 0 \, \, .  \label{scrm}  \eeqa
These relations assure that the pole-terms and the $2\pi$--continuum
do only contribute to subleading orders as $|t|$ becomes large. In
Appendix~\ref{app:A},
we discuss how all these constraints are technically implemented.

Finally, let us count the number of free parameters for the fit
functions eqs.(\ref{ffism},\ref{ffivm}). For $n_s = 3$ isoscalar and $n_v = 3$
isovector mesons, one gets $2(n_s + n_v)= 12$ pole
and $(n_s + n_v) = 6$ mass parameters plus the
pQCD parameter $\Lambda$. We remark here that we do not treat $Q_0$ as
a free parameter since only the combination
$ L^{-1} (0) = [\ln(\Lambda^2 / Q_0^2)]^\gamma$ enters
the pertinent formulae.  We only make sure that $Q_0$ comes out
in the few hundred MeV region because of its proportionality to
$\Lambda_{\rm QCD}$.
We also stress that the $\rho$ contribution does not induce
any free parameter and that the mass of the lowest two isoscalar poles
are fixed to $M_\omega$ and $M_\phi$. With the 12 relations
eqs.(\ref{normm},\ref{neuprom},\ref{scrm}) the effective number of free
parameters is thus $18+1-2 -12 = 5$ (or $6$ if we only
enforce the condition on the neutron charge radius).

\subsection{Additive parametrization}
\label{sec:addfit}
The form factor representation eqs.(\ref{ffism},\ref{ffivm}) is very
simple but one can not directly separate the hadronic from the pQCD
contribution in the spectral functions. Therefore, we present here
another form which is similar in spirit to the one proposed in
ref.\cite{fuwa1}. Under some approximations, one can in fact
construct a spectral representation for the pQCD part which can be
added to the contributions from the meson poles and the two-pion
continuum. This in general
introduces some additional fit parameters $c_i^I$ which are
related to the normalization of the form factors at large $t$,
\beq
F_i^I (t) \to \dfrac{c_i^I}{t^{i+1} \, \bigl[ \ln(-t/Q_0^2) \bigr]^\gamma}
\, \, . \label{fci} \eeq
In addition, one has to take special care about the
singularity at $t_{\rm sing}$ as detailed in appendix~\ref{app:B}.
It leads to an extra pole term not considered in \cite{fuwa1}
\cite{fuwa2}. The explicit additive parametrization takes the form
\beqa
{\rm Re}F^I_i (t) = \dfrac{\delta_{I,(v)}}{\pi}
\int_{4M_\pi^2}^{50M_\pi^2} dt' \dfrac{{\rm Im}F_i^\rho (t')}{t' - t}
+ \sum_I \dfrac{a_i^I}{M^2_I - t} + c_i^I \bar{F}_i^{I, {\rm QCD}} \,
\, \nonumber \\
\bar{F}_i^{I, {\rm QCD}} = \dfrac{\tilde{a}_i^{\rm
QCD} }{\Lambda^2-Q_0^2-t} + \dfrac{1}{\pi} \int_{\Lambda^2}^\infty dt'
\dfrac{{\rm Im}\bar{F}_i^{\rm QCD} (t')}{t' - t} \, \, , \nonumber \\
{\rm Im}\bar{F}_i^{\rm QCD} (t) = \dfrac{ \sin \biggl( \gamma \biggl[
  \arctan \bigl( \dfrac{\pi}{\ln(z)} \bigr) + \pi \Theta\bigl( 1 -
 z \bigr) \biggr] \biggr)}{ \biggl( \bigl( \ln(z)
 \bigr)^2 + \pi^2 \biggr)^{\gamma / 2} \, \, t^{i+1} } \, \, ,
\label{fita}
\eeqa
with $z = |(\Lambda^2 - t) / Q_0^2|$. Clearly, the representation for
the Im~$F_i^{\rm QCD}$ is more complicated than the one used in
\cite{fuwa1} \cite{fuwa2} but it leads to the desired behaviour,
eq.(\ref{fci}). The  normalization and radius
constraints as well as the superconvergence relations are modified,
\beqa
\rho_i^{(-1)} \, \delta_{I, (v)} + \sum_I \,
\dfrac{a_i^I}{M^2_I} + \dfrac{a_i^{I, {\rm QCD}} }{\Lambda^2 - Q_0^2} +
c_i^I \, \xi_i^{(-1)} = \nu_i^I  \, \,
\nonumber \\
\rho_i^{(0)} \, \delta_{I, (v)} + \sum_I \, a_i^I + a_i^{I, \rm QCD} +
c_i^I \, \xi_i^{(0)}
=0 \, \, ,\nonumber \\
\rho_2^{(1)} \, \delta_{I, (v)} + \sum_I \, M^2_I \, a_2^I +
(\Lambda^2 - Q_0^2) \, a_2^{I,{\rm QCD}} +
c_2^I \, \xi_2^{(1)}
= 0 \, \, , \nonumber \\
\dfrac{a_1^{I,{\rm QCD}}}{(\Lambda^2-Q_0^2)^2}
+c_1^I \, \xi_1^{(-2)} + \sum_I \dfrac{a_1^I}{M_I^4}+ \delta_{I,(v)}
\, \rho_1^{(-2)}
=  \dfrac{dF_1^I}{dt}(0)  \, \, \, ,
\label{addcon}  \eeqa
with
\beq
\xi_i^{(k)} = \dfrac{1}{\pi} \int_{\Lambda^2}^\infty dt' \, {t'}^k \,
{\rm Im}F_i^{\rm QCD} (t') \, \, . \eeq
Here, the $\rho^{(k)}_i$ are defined as in eq.(\ref{rho}) with $L (t)
\equiv 1$, and the $a_i^{I, {\rm QCD}}$ are defined in appendix~\ref{app:B}.
The pertinent numerical values are $dF_1^{(s)}(0)/dt =1.37 \,{\rm
  GeV}^{-2}$ and $dF_1^{(v)}(0)/dt =1.30 \,{\rm  GeV}^{-2}$. These last
two conditions in eq.(\ref{addcon}) have to be combined appropriately
if only the neutron charge radius as measured in low--energy
neutron--atom scattering is enforced (i.e. only one constraint results).
In the additive parametrization, eq.(\ref{fita}), we have in principle
three\footnote{Since we take the value of $\Lambda^2$ as given from
the multiplicative parametrization, only three of the $c_i^I$ are independent.}
more effective free parameters than in the multiplicative one.
In the latter case, the normalization of the form factors at large
momentum transfer are essentially fixed by our choice of the functions
$L(t)$.  We stress here that the values one
obtains for the parameters $c_i^I$ are only indicative of the strength
of the form factors in the asymptotic region since their numerical values are
very sensitive to the number of meson poles at low energies one
accounts  for (see also \cite{fuwa1} \cite{fuwa2}). In fact, what can
happen in the additive parametrization is the following. If one does
not give reasonable bounds on the new parameters $c_i^I$, a simple
$\chi^2$--minimization can lead to unphysical results with very large
$|c_i^I |$. These will then influence the low--$t$ behaviour of the
various form factors, in particular the isoscalar radii. Therefore, we
constrain the additive parametrization to essentially give the same low
momentum description of the four em form factors as does the
multiplicative one. This also means that the $c_i^I$ are
fixed, i.e. they are no longer free parameters (see also app.\ref{app:B}).
This method ensures that we can make sensible
statements about the pQCD contributions to the various spectral functions.
Only if one would have data at larger $Q^2$ it would make sense  leaving
the $c_i^I$ as free parameters.

\subsection{Fits with an effective $\rho$ pole}
\label{sec:rhofit}
The inclusion of the $\rho$ contribution as detailed in section \ref{sec:exuni}
leads to a large value for the tensor--vector coupling ratio (as
defined in eq.(\ref{kappa})) \cite{hopipi},
\beq \kappa_\rho = 6.6 \pm 1 \, \, \, . \eeq
This value is in agreement with other determinations,
see e.g. Grein \cite{grein} ($\kappa_\rho = 6.1 \pm 0.6$).
More recently,
Brown and Machleidt \cite{brown} have discussed the evidence
for a strong $\rho NN$ coupling from the
measurements of the $\epsilon_1$ parameter in $NN$ scattering.
In ref.\cite{fuwa1} an effective $\rho$
pole with a mass of $0.63$ GeV was used and led to typical values of
$\kappa_\rho = 5.9$. We will also perform such a simple pole fit, i.e.
substituting the full two--pion continuum by a $\rho$ pole with a
variable mass and taking in addition two more pole terms in the
isovector channels. Our motivation to perform such types of simplified
fits is to check whether the large value of the
tensor--to--vector coupling of the $\rho$ with the correct
implementation of the pQCD constraints can be considered a generic
result.

\section{Results and discussion}
\label{sec:res}
Before discussing the specific results of our fits, we wish to make
some general comments. We had argued before that the masses of the
three isovector excitations and of the highest isoscalar one need not
to coincide with masses of physical particles.
However, we have found that it is possible to find a minimum
in the $\chi^2$ hyper--surface
fixing $M_{\rho'} = 1.45$~GeV and $M_{\rho''} = 1.69$~GeV together with
$M_{S'} = 1.60$~GeV. These are the values of the most recent particle
data group compilation \cite{pdg} for the lowest
isovector--vector and isoscalar--vector
meson excitations. Leaving the values of these masses completely free
does not alter the $\chi^2$ significantly.
In contrast, the mass of the third  pole in the
isovector channel is tightly bound due to the various
constraints the fits have to obey. We observe that the mass of this third
isovector pole tends to come  out close to one of the other isovector poles,
thus an effective double--pole structure around $M_{\rho'}$
or $M_{\rho''}$ emerges. This is in marked contrast to the findings of
ref.\cite{hoeh76}. However, it is mandatory to retain three poles
besides the two--pion cut contribution in the isovector channel.
Also, independent of the details of the fits, we
find that while the form factors $F_1^{(s)}(t)$ and
$F_2^{(v)}(t)$ exhibit a stable
dipole structure (i.e. the lowest two poles have residua which are
equal in magntiude but with different signs), this is not the case for
$F_2^{(s)}(t)$ and $F_1^{(v)}(t)$. These findings agree with the ones of
\cite{hoeh76}. Concerning the accuracy of our fits, all normalization,
radius and superconvergence relations are fulfilled within machine
accuracy, typically much better than one part in $10^{12}$.
After these general remarks, we turn to a more detailed
description of our results.

\subsection{The best fit: Form factors, radii and coupling constants}
\label{sec:best}
The optimal fit to the available set of form factor data is obtained
with the isovector masses of 1.45, 1.65 and 1.69 GeV, respectively and
the isoscalar ones being 0.782, 1.019 and 1.60 GeV. The corresponding
residua are $a_1^{\rho'} = -3.465$, $a_2^{\rho'} = -6.552$,
$a_1^{\rho''} = 40.26$, $a_2^{\rho''} = 7.881$, $a_1^{\rho'''} =
-37.30$, $a_2^{\rho'''} = -2.821$, $a_1^\omega = 0.747$,
$a_2^\omega = -0.122$, $a_1^\Phi = -0.738$, $a_2^\Phi = 0.162$,
$a_1^{S'} = -0.0382$ and $a_2^{S'} = -0.0406$.
We find $\Lambda^2 = 9.73$~GeV$^2$
and $Q_0^2 = 0.35$~GeV$^2$. We notice that to have good fits, we can
vary $\Lambda^2$ between 5 and 15~GeV$^2$ without drastically changing
any of our conclusions.
We always work with $\gamma = 2.148$ since the
fits are completely insensitive to the possible variation in this quantity.
In this fit, only the constraint on the neutron charge radius is
imposed and the $\rho-\omega$ mixing in the two--pion spectral
function is included. The resulting $\chi^2$/datum is 1.09.

In Fig.4, we show the electric and magnetic proton and neutron form
factors normalised to the dipole fit.\footnote{In the case of $G_E^n$,
 we divide by $G_E^n$ as given by Platchkov et al. \cite{platch}
(denoted by $G_P$) since in the dipole approximation, $G_E^n \equiv
 0.$} Similar to earlier findings, we note that there are substantial
deviations from the dipole fit in all channels, particularly at large
momentum transfer. We also note that a better data basis is clearly
needed.

Of particular importance is the determination of the nucleon radii.
In table~1, we give radii corrsponding to the Pauli and Dirac form
factors in comparison to the results of ref.\cite{hoeh76}. For the
isovector form factors,  the radii are indeed dominated
by the two--pion plus $\rho$ contribution, we have $r_1^\rho  \simeq
r_2^\rho \simeq 0.75$~fm. The
corresponding neutron and proton radii are given in table~2. The
uncertainty for these radii is 1\% (for comparison, in \cite{hoeh76} the
uncertainties on the radii were of the order of 3\%). This number
is calculated in the following way. In the
parameter--space we look for solutions with a comparable
$\chi^2$/datum than the best fit has. Equivalently, one can sum in
quadrature the $1\sigma$ deviations of these parameters contributing
to the various radii. Our results are comparable to the ones
of ref.\cite{hoeh76} with the exception of $r_2^{(v)}$ and
$r_M^n$. This can be traced back to the fact that in \cite{hoeh76} the
superconvergence relation
\beq
\dfrac{1}{2} a_2^\rho (d_2^\rho)^2 +
\sum_{V = \rho' , \rho'', \rho'''} \,a_2^V \, M^2_{V} = 0 \, \, ,
\label{sn}
\eeq
was not taken into account. In contrast, our values for
$r_2^{(v)}$ and $r_M^n$ are based on a completely consistent calculation.
We point out that there exist some on--going activity e.g. at ELSA (Bonn)
to determine the neutron magnetic form factor more precisely at low
and moderate momentum transfer.
We also note that the value for $r_E^p$ is on the low side of the result of
ref.\cite{simon}. If one insists on their central value, $r_E^p =
0.862$~fm, by imposing the proper slope condition, one is not able to
simultaneously fit the neutron and the proton form factor data. In
fact, in \cite{simon} only proton data were considered. We conclude
that the uncertainty attributed to $r_E^p$ in \cite{simon} is
presumably underestimated. As we already anticipated in section
\ref{sec:prot}, the
constraint from the proton charge radius in its present form is too
restrictive to be applied in the dispersive analysis. Better
low--energy data are clearly called for to settle this issue.

The mesonic coupling constants can be directly inferred from the
pertinent residua. Applying the same analysis to get a handle at the
uncertainties  as described above, we find
\beqa
g_1^{\omega NN} = 20.86 \pm 0.25 \, \, ,
\quad g_2^{\omega NN} = -3.41 \pm 0.24 \, \, ,
\nonumber \\
g_1^{\Phi NN} = -9.16 \pm 0.23 \, \, ,
\quad g_2^{\Phi NN} = 2.01 \pm 0.33 \, \, ,
\label{gcoup}
\eeqa
which are compared with the findings of ref.\cite{hoeh76} in
table~3. Our $\Phi$--couplings are somewhat larger but consistent
within error bars. We note that the $\omega NN$ coupling is larger
than in typical one--boson exchange potentials or from the analysis of
Grein and Kroll \cite{greink} using forward dispersion relations
 for $NN$--scattering,
$(g_1^{\omega NN})^2 / 4\pi= 8.1 \pm 1.5$. Such a small coupling
constant value cannot be accomodated in our fit, it is inconsistent
with empirical information on the slope of $F_1^{(s)}$ if the
$\omega$ and the $\Phi$ lead to the dipole structure as described
above. This point is also
discussed in some detail in \cite{ho93}.
Furthermore, we remark that a direct comparison with coupling
constants obtained in  boson--exchange models, which in general
include strong meson-nucleon form factors, has to be taken cum grano
salis.\footnote{The photon couplings through vector mesons to the nucleon
need not be the same than the purely strong interaction $VNN$ couplings}.
In flavor SU(3), one can
derive the following formulae for the $\omega - \Phi$--mixing angle, $\Theta$,
\beq
\dfrac{\sqrt{3}}{\cos(\Theta)}\dfrac{g_1^{\rho NN}}{g_1^{\omega NN}} -
\tan(\Theta) = \dfrac{g_1^{\Phi NN}}{g_1^{\omega NN}} \, \, . \label{mix}
\eeq
Using $g_1^{\rho NN} = 2.6$ \cite{hopipi}, we have $\Theta =
35^\circ$, very close to the ideal mixing angle of $37^\circ$. This
means that the $\Phi$ is almost entirely an $\bar s s $ state and is
thus supposed to decouple from the nucleon
(to leading order in flavor perturbation theory). This is the much discussed
violation of the OZI rule. This apparent paradox finds its resolution
in the fact that the simple pole approach for the $\Phi NN$ coupling
effectively includes contributions from the $\bar K K$
continuum. Stated differently, the $\Phi$ can couple to the kaon cloud
surrounding the nucleon (as modeled e.g. in \cite{gakr3}). This topic
could be investigated further along the lines discussed in section
\ref{sec:exuni}, i.e. by analyzing the $\bar N N \to \bar K K$ partial waves.

We close the discussion about the coupling constants with some remarks
on $\kappa_\rho$. As argued in section \ref{sec:rhofit}, one can also
perform fits with a $\rho$--pole. In this case, one cannot take
the physical mass for the $\rho$
since otherwise the isovector radii are severely underestimated. We
have performed such fits and find
\beq
\kappa_\rho = 6.1 \pm 0.2 \, \, ,
\eeq
which is consistent with previous determinations as discussed in
section \ref{sec:rhofit}. We consider this an important consistency
check on our fits. The realistic fits, however, have to include the
correlated two--pion exchange as described in section \ref{sec:exuni}.

Next, we discuss the large momentum behaviour of the form factors.
Only for $G_M^p$, there are data for $Q^2 > 10$~GeV$^2$. In Fig.5,
we show the quantities $Q^4 G_M^p(Q^2)/\mu_p$ and
$L^{-1}(Q^2) Q^4 G_M^p(Q^2)/(\mu_p L^{-1} (0) )$
up to $Q^2= 50$~GeV$^2$. Within the uncertainties, the curve
representing the second function
tends to a constant (as it is expected from pQCD), but it is obvious
that more precise data at high momentum transfer are mandatory to
really pin down this behaviour. The available data do not exclude that
asymptotia sets in much beyond $Q^2 = 30$~GeV$^2$. Often considered is
also the ratio $Q^2 F_2 (Q^2)/ F_1 (Q^2)$ which should become constant
as $Q^2$ becomes large (the extra $Q^2$ in front of $F_2$ compensates
the spin--flip suppression $\sim 1/Q^2$). In this ratio, many
uncertainties related to the exact form of the nucleon wave function
drop out. As seen in Fig.6, the presently available data are at too
low momentum transfer to test this prediction, although there is some
hint of $Q^2 F_2 (Q^2)/ F_1 (Q^2)$ becoming constant as $Q^2$
increases. Again, more accurate data at higher $Q^2$ are called for.

\subsection{Additive parametrization: Hadron versus quark
contributions}
\label{sec:addres}
As explained before, the multiplicative parametrization is not
well suited for separating the hadronic (pole) from the quark (pQCD)
contributions. That is the reason underlying the additive
parametrization. In this case, the four normalization constants
$c_i^I$ (of which only three are independent) in principle
increase the number
of tuneable parameters. However, as explained before, one has to
impose certain constraints on the actual values of the $c_i^I$ since
otherwise completely unphysical solutions of the fitting procedure can
emerge. Our strategy is therefore to constrain pole parameters and the
$c_i^I$ such
that we essentially recover the low momentum description of the
multiplicative parametrization, in particular the nucleon radii. Only
with such constraints one can make sensible statements about the
separation of hadronic and pQCD contributions.

In Fig.7, we show a typical result for the isoscalar and isovector
form factors. We have $c_1^{(v)} = -53.40$, $c_1^{(s)} = -8.20$,
$c_2^{(v)} = 95.56$ and $c_2^{(s)} = -1.55$. The $\chi^2$/datum is
1.86. This increased value is a mostly reflection of the the
approximations performed (i.e. the neglect of subleading $1/t$
corrections) to derive the additive parametrization, compare
app.\ref{app:B}.\footnote{In that appendix, it is also shown how one
can systematically improve this procedure. For our purpose, the lowest order
approximation used here is, however, sufficient.}
We repeat that leaving all parameters
free, one could naturally find a solution with a lower $\chi^2$ than
for the multiplicative parametrizations. Such solutions, however, have
to be discarded as discusssed before.
For all form factors, the hadronic and the quark contribution are of
comparable size (in magnitude) around $\Lambda^2 \simeq 10$~GeV$^2$.
We notice that for the Dirac
form factors $F_{1}^{(s,v)} (Q^2)$,
the hadronic contribution quickly drops off beyond
$Q^2$ larger than 10~GeV$^2$. For the Pauli form factors
$F_{2}^{(s,v)} (Q^2)$, this fall--off is slower which essentially
is the reason that one does not observe pQCD scaling for the available
data.  In all cases, the quark contribution is very small at low $t$
(by construction). A similar behaviour was
noticed in \cite{fuwa1}, \cite{fuwa2} although
in these papers the asymptotic behaviour was incorrectly implemented
in the spectral functions.
Of course, with the presently available data base, these results
should only be considered indicative. With better data in the few
and many GeV region, one will eventually be able to more cleanly
separate the hadronic from the quark contribution. In particular, for
the range of momentum transfer available at CEBAF, one will
essentially probe the transition region from the hadronic to the quark
description. The planned experiments at CEBAF \cite{cebaf} will
certainly shed light on this interesting regime.

\vspace{3cm}

\section*{Acknowledgements}
We thank Gerhard H\"ohler for many informative discussions and Ralf Gothe
for providing us with the most recent ELSA data. We are also grateful
to Wulf Kr\"umpelmann for supplying us with his data basis.

\newpage

\begin{appendix}
\section{Implementation of the constraints}
\label{app:A}
In this appendix we show how to evaluate the fit constraints. In general we
have a system of 12 equations. It can be written in the following way,
\beq
\hat{M}_{i}^{I} \vec{A}_{i}^{I} = \vec{C}_{i}^{I}
\quad,
\eeq
where the matrix $\hat{M}_{i}^{I}$ has the form
\beq
\hat{M}_{i}^{I} =
\left( \begin{array}{ccc}
M_{I_{1}}^{-2} & M_{I_{2}}^{-2} & M_{I_{3}}^{-2} \\
1 & 1 & 1 \\
\tilde{M}_{I_{1}} & \tilde{M}_{I_{2}} & \tilde{M}_{I_{3}}
\end{array} \right) \quad,\quad
\tilde{M}_{I} = \sigma \cdot \left\{ \begin{array}{ll}
\frac{1}{M_{I}^{4}}+\kappa \frac{\gamma L^{1/\gamma}(0)}{M_{I}^{2}
\Lambda^2} &,\quad i=1 \\
M_{I}^{2} &,\quad i=2 \\
\end{array} \right. \, \, .
\eeq
Setting $\sigma=1$, $0$, respectively, permits to switch off certain vector
meson pole terms. $\kappa$ specifies the parametrization.
$\kappa=1$ if the multiplicative parametrization is used and
$\kappa=0$ for the additive one. With $A_{i}^{I} =
a_{i}^{I}L^{-1}(M_{I}^2)$ we can write
\beq
\vec{A}_{i}^{I} = \left(\begin{array}{c}
A_{i}^{I_{1}} \\ A_{i}^{I_{2}} \\ A_{i}^{I_{3}}
\end{array} \right) \quad,\quad
\vec{C}_{i}^{I} = \left(\begin{array}{c}
\nu_{i}^{I} L^{-1}(0) - \rho_{i}^{(-1)} \delta_{_{I,(v)}} \\
-\rho_{i}^{(0)} \delta_{_{I,(v)}} \\
\tilde{C}_{i}^{I}
\end{array} \right) \quad.
\eeq
The vector $\vec{C}_{i}^{I}$ contains all fit constraints. The first
component enforces the correct normalizations, the second and the last one
in the case ($i=2$) induce the superconvergence relation (\ref{scrm}). For
($i=1$) the third component implies the slope informations stemming from the
experimental values of the electric radii of proton and neutron,
\beq
\tilde{C}_{i}^{I} = \left\{\begin{array}{ll}
\frac{dF_{1}^{I}}{dt}(0)L^{-1}(0)- \Re \delta_{_{I,(v)}} &,\quad i=1 \\
-\rho_{2}^{(1)} \delta_{_{I,(v)}} &,\quad i=2
\end{array} \right. \quad,
\eeq
where $\Re$ is the contribution of the two pion continuum to the radius
constraints
\beq
\Re = \left(a_{1}^{\rho}L^{-1}(M_{a}^{2}) + b_{1}^{\rho}L^{-1}(M_{b}^{2})
\right)\left(\frac{1}{d_{1}^{\rho}}+\kappa\frac{\gamma L^{1/\gamma}(0)}
{2 \Lambda^2} \right) + 2 \frac{b_{1}^{\rho}L^{-1}(M_{b}^{2})}{c_{1}^{\rho}}
\quad.
\eeq
We simply invert the matrix $\hat{M}_{i}^{I}$ to find the solution vector
$\vec{A}_{i}^{I}$,
\beq
(\hat{M}_{i}^{I})^{-1} = \frac{1}{m}
\left( \begin{array}{ccc}
\tilde{M}_{I_{3}}-\tilde{M}_{I_{2}} &
M_{I_{3}}^{-2}\tilde{M}_{I_{2}}-M_{I_{2}}^{-2}\tilde{M}_{I_{3}} &
M_{I_{2}}^{-2}-M_{I_{3}}^{-2} \\
\tilde{M}_{I_{1}}-\tilde{M}_{I_{3}} &
M_{I_{1}}^{-2}\tilde{M}_{I_{3}}-M_{I_{3}}^{-2}\tilde{M}_{I_{1}} &
M_{I_{3}}^{-2}-M_{I_{1}}^{-2} \\
\tilde{M}_{I_{2}}-\tilde{M}_{I_{1}} &
M_{I_{2}}^{-2}\tilde{M}_{I_{1}}-M_{I_{1}}^{-2}\tilde{M}_{I_{2}} &
M_{I_{1}}^{-2}-M_{I_{2}}^{-2}
\end{array} \right) \quad,\quad
\eeq
with
\beq
m = (M_{I_{2}}^{-2}-M_{I_{3}}^{-2})\tilde{M}_{I_{1}}+
(M_{I_{3}}^{-2}-M_{I_{1}}^{-2})\tilde{M}_{I_{2}}+
(M_{I_{1}}^{-2}-M_{I_{2}}^{-2})\tilde{M}_{I_{3}}
\quad.
\eeq

\newpage

\section{Derivation of the additive parametrization}
\label{app:B}

Here, we derive the additive parametrization, eq.(\ref{fita}). The
Cauchy integral representation  for the fit
functions $F_i^I (t)$ ($i=1,2; I=(v),(s)$),
eqs.(\ref{ffism},\ref{ffivm}), takes the form
\beq
F_i^I (t) = \dfrac{1}{2\pi i}\oint_{{\cal C}_I} dt'
\dfrac{F_i^I (t')}{t' - t} \, \, ,
\eeq
with ${\cal C}_I$ a closed integration contour as shown in Figs.3a,b.
${\cal C}_I$ is chosen such that Re~$F_i^I (t)$ can be calculated as a
Hilbert-transform. For that, we need the imaginary part of $F_i^I
(t)$,
\beq
{\rm Im}~F_i^I (t \pm i \epsilon ) = {\rm Re}~{\tilde F}_i^I (t) \,
{\rm Im}~L ( t \pm i \epsilon ) + {\rm Im}~{\tilde F}^I_i (t \pm i \epsilon )
\, {\rm Re}~L(t) \, \, .
\eeq
The real and imaginary parts of $F_i^I (t)$ for $t< t_{\rm sing} =
\Lambda^2 - Q_0^2$ and $t > t_{\rm sing}$ can be deduced
easily. $L(t)$ is purely real for $t < t_{\rm sing}$. Consider now the
imaginary part of $L(t)$. For $t_{\rm
  sing} < t < \Lambda^2$, we find (on the first Riemann sheet)
\beq
{\rm Im}~L(t \pm i \epsilon ) = \mp \dfrac{\sin(\gamma \pi)}{|\ln(z)|^\gamma}
\, \, ,  \eeq
with
\beq   z = \biggl| \dfrac{\Lambda^2  - t}{Q_0^2} \biggr| \, \, , \eeq
and for $t>\Lambda^2$, it reads (on the first sheet)
\beq
{\rm Im}~L (t \pm  i \epsilon) = \mp\dfrac{ \sin \biggl( \gamma \biggl[
  \arctan \bigl( \dfrac{\pi}{\ln(z)} \bigr) + \pi \Theta\bigl( 1 -
 z \bigr) \biggr] \biggr)}{ \biggl( \bigl( \ln(z)
 \bigr)^2 + \pi^2 \biggr)^{\gamma / 2} } \, \, ,
\eeq
and finally
\beq
{\rm Im}~L (\Lambda^2 \pm i \epsilon ) = 0 \, \, , \eeq
assuming that $M^2_I < \Lambda^2$ for all $I$. Due to the singularity
at $t_{\rm sing}$, special care has to be taken of the region $t
\in [\Lambda^2 - 2Q_0^2, \Lambda^2]$. We have
\beq
{\rm Re}~F_i^I (t) = \dfrac{1}{\pi} \int_{\Lambda^2}^\infty dt'
\dfrac{{\rm Re}~\tilde{F}_i^I (t') \,{\rm Im}~L(t)}{t' -t} +
\dfrac{1}{2\pi} {\rm Im} \, \biggl[ \oint_{{\cal C}_I^{\rm sing}} dt'
\dfrac{F_i^I (t') }{ t' - t } \biggr] + \dots\, \, , \eeq
where the ellipsis stands for the contributions from the pole terms and
the two--pion continuum (which are of no relevance for the large--t
behaviour discussed here). For the contour ${\cal C}_I^{\rm sing}$ we
choose a closed circle around the singularity with radius $Q_0^2$.
In the region of integration and close to the singularity, we have
$t \gg M_I^2$ so that we can approximate $\tilde{F}_i^I (t)$ by its
asymptotic behaviour
\footnote{In our fits, the most massive isovector (isoscalar)
pole is at 2.8~(2.6)~GeV$^2$, which is a bit close to the value of $\Lambda^2
\simeq 10$~GeV$^2$. This induces some uncertainty due to the terms
 neglected.}.
\beq
{\rm Re}~F_i^I (t) = \dfrac{c_i^I}{\pi} \int_{\Lambda^2}^\infty dt'
\dfrac{{\rm Im}~L(t)}{(t')^{i+1} \, (t' -t )} +
\dfrac{c_i^I}{2\pi} {\rm Im} \, \biggl[ \oint_{{\cal C}_I^{\rm sing}} dt'
\dfrac{L (t') }{(t')^{i+1} \, (t' - t) } \biggr] + \dots\, \, .
\label{refi} \eeq
Let us concentrate on the second term in
eq.(\ref{refi}). Parametrizing the integration path via $t(\phi) =
t_{\rm sing} + Q_0^2 \exp(i \phi)$, we find
\beq
I(t) = \dfrac{1}{2\pi}\int_0^{2\pi}d\phi \dfrac{ i Q_0^2 e^{i \phi }
  \ln \bigl(1 - e^{i \phi} \bigr)^{-\gamma} }{ \bigl[\Lambda^2 - Q_0^2
  ( 1- e^{i \phi}) \bigr]^{i+1}  \bigl[\Lambda^2 - t -Q_0^2
( 1- e^{i \phi}) \bigr] } \, \, . \label{itap} \eeq
We expand the first term in the numerator in eq.(\ref{itap}),
\beq
\dfrac{1}{\bigl[\Lambda^2 - Q_0^2  ( 1- e^{i \phi}) \bigr]^{i+1}} =
\dfrac{1}{\bigl[\Lambda^2 - Q_0^2 \bigr]^{i+1}}  \, \biggl( 1 - (i+1)
\dfrac{Q_0^2}{\Lambda^2 - Q_0^2} \, e^{i \phi}  + \ldots \biggr) \, \,
, \eeq
and drop the term proprotional to $\exp(i \phi)$ in the numerator
of eq.(\ref{itap}). Altogether, we find
\beq
I(t) = i \, \dfrac{\tilde{a}_I^{\rm QCD} }{\Lambda^2 - Q_0^2 - t} \,
\, , \eeq
with
\beqa
\tilde{a}_I^{\rm QCD}  = \dfrac{Q_0^2}{2\pi (\Lambda^2 -
Q_0^2)^{i+1} } \biggl[ C^{(1)} - (i+1) \dfrac{Q_0^2}{\Lambda^2 -
Q_0^2} \, C^{(2)} + \ldots \biggr] \, \, , \nonumber \\
C^{(n)} = \int_0^{2 \pi} d\phi \, e^{i n\phi } \, \ln (1- e^{i\phi}
)^{-\gamma} \qquad \qquad \, \, ,
\eeqa
where the coefficients $C^{(n)}$ can be found by numerical
integration. They are purely real. It is sufficient to keep the first
two terms in the series with $C^{(1)} = -5.861$ and $C^{(2)} = 7.129$.
Setting now
\beq
a_i^{I, {\rm QCD}} = c_i^I \, \tilde{a}_i^{\rm QCD} \, \, , \eeq
we have arrived at the desired result, eq.(\ref{fita}).

Finally, we wish to establish a more rigorous derivation which is
useful if one wants to retain more terms of the asymptotic expansion
of the ${F}^I_i (t)$. For that, we  expand the $\tilde{F}_i^I (t)$
in powers of $1/t$,
\beq
\tilde{F}_i^I (t) = \sum_{k_{i} \le (i+1)}^{\infty} c_{k_{i}}^{I} t^{-k_{i}}
\, \, ,
\nonumber
\eeq
with the coefficients $c_{k_{i}}^{I}$  given by
\beq
c_{k_{i}}^{I} = (-)^{k_{i}} \left( \sum_{I}
a_{i}^{I} \, M_{I}^{2\,(k_{i}-1)} \, L^{-1}(M_{I}^{2}) +
\frac{\delta_{I,(v)}}{\pi}
\int_{4\,M_{\pi}^2}^{50\,M_{\pi}^2} dt' (t')^{k_{i}-1} {\rm Im}
\,F_{i}^{\rho}(t') L^{-1}(t') \right) \, \, ,
\eeq
and
\beq
c_{k_{i} < (i+1)}^{I} = 0
\eeq
because of the superconvergence relations. This leads to
\beq
{\rm Re}~F_i^I (t) = \sum_{k_{i} \le (i+1)}^{\infty} c_{k_{i}}
\left(
\dfrac{1}{\pi} \int_{\Lambda^2}^\infty dt'
\dfrac{{\rm Im}~L(t)}{(t')^{k_{i}} \, (t' -t )} +
\dfrac{1}{2\pi} {\rm Im} \, \biggl[ \oint_{{\cal C}_I^{\rm sing}} dt'
\dfrac{L (t') }{(t')^{k_{i}} \, (t' - t) } \biggr]
\right)
 + \dots\, \, ,
\eeq
which can be written in a form  similar to eq.(\ref{fita})
\beq
{\rm Re}F^I_i (t) = \dfrac{a_{i}^{I,\rm QCD} }{\Lambda^2-Q_0^2-t} +
\dfrac{1}{\pi} \int_{\Lambda^2}^\infty dt'
\dfrac{{\rm Im}F_i^{\rm QCD} (t')}{t' - t} \,\, +\ldots \, \, ,
\eeq
but in this case we find
\beqa
a_{i}^{I,\rm QCD} &=& \sum_{k_{i} \le (i+1)}^{\infty}
\dfrac{c_{ k_{i} }^{I} Q_0^2}{2\pi (\Lambda^2 -
Q_0^2)^{ k_{i} }} \biggl[ C^{(1)} - k_{i} \dfrac{Q_0^2}{\Lambda^2 -
Q_0^2} \, C^{(2)} + \ldots \biggr] \quad, \nonumber\\
{\rm Im}F_i^{\rm QCD} (t) &=& {\rm Re}\tilde{F}_{i}^{I}(t) \, {\rm Im} L(t)
\, \, \, .
\label{asss}
\eeqa
Considering only the first term of the sum over
$k_{i}$ in eq.(\ref{asss}) is equivalent to the approximation eq.(\ref{refi}).

\section{Parametrization of the best fit}
\label{app:C}
Here we give the parametrization of the best fit discussed in
section \ref{sec:best} explicitly for easy usage, with $Q^2$ in GeV$^2$,
\beqa
F_{1}^{(s)}(Q^2) &=& \left[\frac{9.464}{0.611+Q^2} -
\frac{9.054}{1.039+Q^2} - \frac{0.410}{2.560+Q^2} \right]
\left[\ln\left(\frac{9.733+Q^2}{0.350}\right)\right]^{-2.148}
\quad,
\nonumber\\
F_{2}^{(s)}(Q^2) &=& \left[- \frac{1.549}{0.611+Q^2} +
\frac{1.985}{1.039+Q^2} - \frac{0.436}{2.560+Q^2} \right]
\left[\ln\left(\frac{9.733+Q^2}{0.350}\right)\right]^{-2.148}
\quad,
\nonumber\\
F_{1}^{(v)}(Q^2) &=& \Bigg[\frac{1.032
\left[\ln\left(\frac{9.733-0.500}{0.350}\right)\right]^{2.148}
+0.088\left[\ln\left(\frac{9.733-0.400}{0.35}\right)\right]^{2.148}
\left(1+\frac{Q^2}{0.318}\right)^{-2}}{2\left(1+\frac{Q^2}{0.550}\right)}
-
\nonumber\\
& &\quad \frac{38.885}{2.103+Q^2} +  \frac{425.007}{2.734+Q^2}
- \frac{389.742}{2.835+Q^2} \Bigg]
\left[\ln\left(\frac{9.733+Q^2}{0.350}\right)\right]^{-2.148}
\quad,
\nonumber\\
F_{2}^{(v)}(Q^2) &=& \Bigg[\frac{5.782
\left[\ln\left(\frac{9.733-0.500}{0.350}\right)\right]^{2.148}
+0.391\left[\ln\left(\frac{9.733-0.400}{0.350}\right)\right]^{2.148}
\left(1+\frac{Q^2}{0.142}\right)^{-1}}{2\left(1+\frac{Q^2}{0.536}\right)}
-
\nonumber\\
& &\quad \frac{73.535}{2.103+Q^2} +  \frac{83.211}{2.734+Q^2}
- \frac{29.467}{2.835+Q^2} \Bigg]
\left[\ln\left(\frac{9.733+Q^2}{0.350}\right)\right]^{-2.148}
\quad.
\eeqa

\end{appendix}

\newpage

\centerline{\Large {\bf Tables}}

\vspace{2cm}

\begin{center}

\renewcommand{\arraystretch}{1.5}

\begin{tabular}{|c||c|c|c|c|} \hline
& $r_1^{(s)}$ [fm]& $r_2^{(s)}$ [fm] & $r_1^{(v)}$ [fm]
& $r_2^{(v)}$ [fm]   \\ \hline
Best Fit          & 0.782  & 0.845 & 0.765 & 0.893 \\
Ref.\cite{hoeh76} & 0.77   & 0.837 & 0.76  & 0.863 \\
\hline \end{tabular}

\bigskip

Table 1: Radii of the Dirac and Pauli form factors.

\end{center}

\vspace{2cm}

\begin{center}

\renewcommand{\arraystretch}{1.5}


\begin{tabular}{|c||c|c|c|c|c|c|} \hline
& $r_E^p$ [fm]& $r_M^p$ [fm] & $r_M^n$ [fm]
& $r_1^p$ [fm]& $r_2^p$ [fm] & $r_2^n$ [fm]
\\ \hline
Best Fit          & 0.847  & 0.836 & 0.889 & 0.774 & 0.894 & 0.893 \\
Ref.\cite{hoeh76} & 0.836  & 0.843 & 0.840 & 0.761 & 0.883 & 0.876 \\
\hline \end{tabular}

\bigskip

Table 2: Proton and neutron radii.

\end{center}

\vspace{2cm}

\begin{center}

\renewcommand{\arraystretch}{1.5}

\begin{tabular}{|c||c|c|c|c|} \hline
& $(g_1^{\omega NN})^2/4\pi$ & $\kappa_\omega$
& $(g_1^{\Phi NN})^2/4\pi$   & $\kappa_\Phi$    \\ \hline
Best Fit          & $\, 34.6 \pm 0.8 \, $  & $\, -0.16 \pm 0.01 \,$
& $\, 6.7 \pm 0.3 \,$ & $ \, -0.22 \pm 0.01 \, $  \\
Ref.\cite{hoeh76} & $30 \pm 3$   & $-0.17$ & $4.4 \pm 1$  & $-0.3$ \\
\hline \end{tabular}

\bigskip

Table 3: Coupling constants of the isoscalar vector mesons.

\end{center}

\newpage

\centerline{\Large {\bf Figure captions}}

\medskip

\begin{enumerate}
\item[Fig.1] Two--pion cut contribution to the isovector nucleon form
  factors.

\item[Fig.2] Spectral distribution of the  isovector form
factors weighted with $1/t^2$. Shown are 2Im$F_1^v (t)/t^2$
(long--dashed line),
2$\tau$Im$F_2^v (t)/t^2$ (short--dashed line) and
2Im$G_E^v (t)/t^2$ (solid line). Upper panel: no $\rho$--$\omega$
mixing. Lower panel: with $\rho$--$\omega$ mixing.

\item[Fig.3] Analytic structure of the fit functions
eqs.(\ref{ffism},\ref{ffivm}) in the complex $t$ plane. (a)
Isoscalar case. (1) Poles at $M_\omega$, $M_\Phi$ and $M_{S'}$,
(2) the singularity at
$\Lambda^2 - Q_0^2$ and (3) the right--hand cut starting at
$\Lambda^2$. The data are given at negative $t$. (b)
Isovector case. The right--hand cut (1) is the two--pion continuum,
(2) are the three isovector poles,
(3) and (4) are equivalent to (2) and (3) of the isoscalar case.

\item[Fig.4] Optimal fit within the multiplicative parametrization
including $\rho$--$\omega$ mixing and the constraint from the neutron
charge radius. The data for $G_E^n$ are from \cite{galster}
\cite{albre1} \cite{rock}
\cite{lung} \cite{platch} \cite{wood} \cite{eden} \cite{mey},
for $G_E^p$  from
\cite{berger} \cite{bartel} \cite{albre2} \cite{jans} \cite{hoeh76}
\cite{bork} \cite{simon} \cite{simon2} \cite{walker} \cite{bosted},
for $G_M^p$ from
\cite{berger} \cite{bartel} \cite{jans} \cite{hoeh76}
\cite{bork}  \cite{simon2} \cite{arnold} \cite{walker2},
and for $G_M^n$ from \cite{stein} \cite{hughes} \cite{aker} \cite{dunn}
\cite{bonn}. For the data of ref.\cite{platch}, we have taken the
values based on the Paris potential but enlarged the error bars to
account for the model--dependence.

\item[Fig.5] Asymptotic behaviour of the proton magnetic form factor.
The best fit in comparison to the available data is shown.
Upper panel: $Q^4 G_M^P (Q^2) /\mu_p$, lower panel:
$L^{-1}(Q^2) Q^4 G_M^P (Q^2) /( \mu_p L^{-1}(0) )$

\item[Fig.6] The ratio $Q^2 F_2^p (Q^2) / F_1^p (Q^2)$
for the best fit compared to the data.

\item[Fig.7] Hadronic (short--dashed lines) versus quark (long--dashed
lines) contributions for the isoscalar and isovector Dirac and Pauli
form factors (solid lines) in the additive  parametrization. The
absolute values of the various contributions are shown.

\end{enumerate}

\newpage

\begin{figure}[bht]

$\;$\vspace{2cm}

\centerline{
\epsfxsize=2.5in
\epsfysize=1.5in
\epsffile{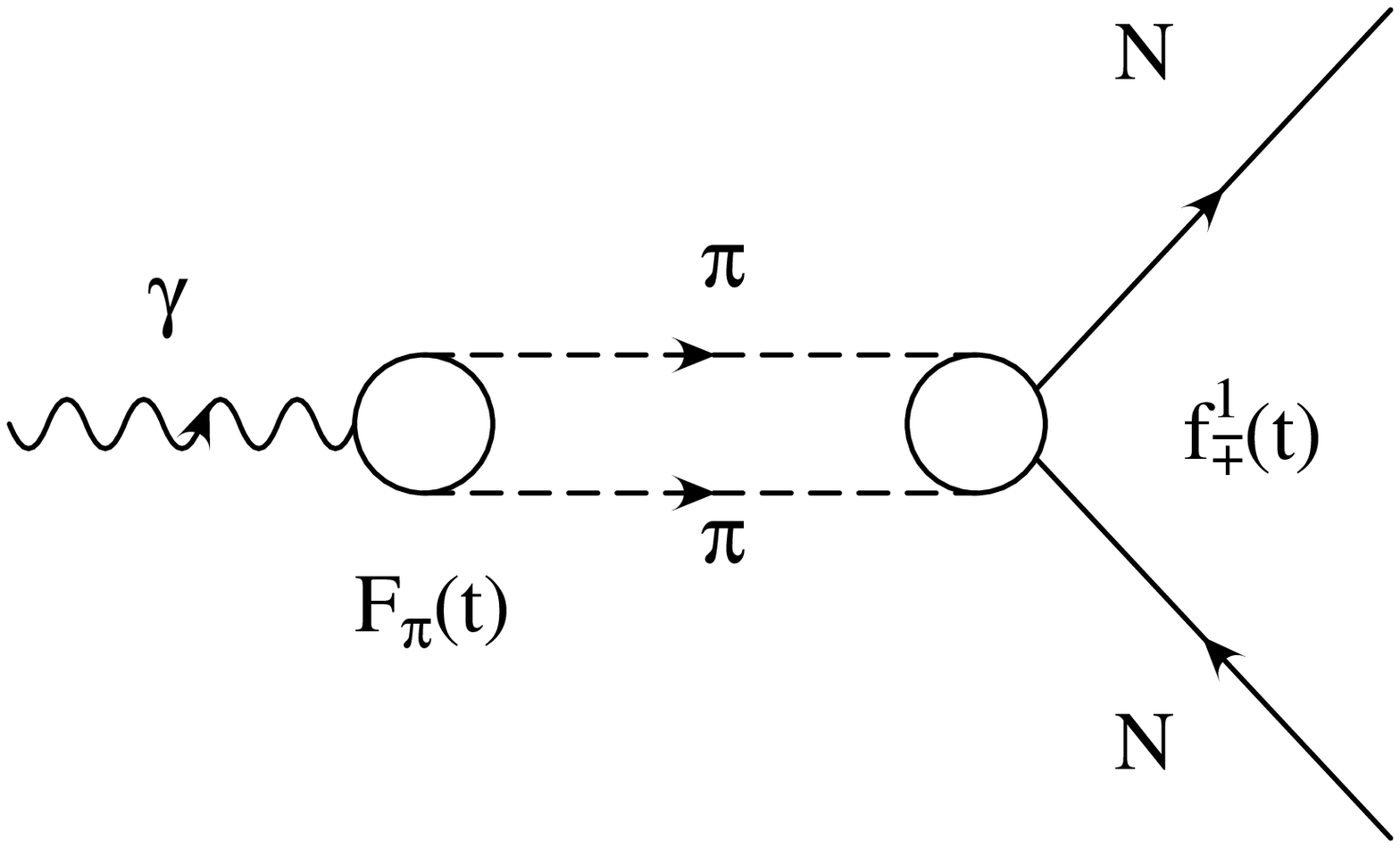}
}
\vskip 0.7cm

\centerline{\Large Figure 1}
\end{figure}

$\;$\vspace{1cm}

\begin{figure}[bht]
\centerline{
\epsfxsize=3in
\epsfysize=4in
\epsffile{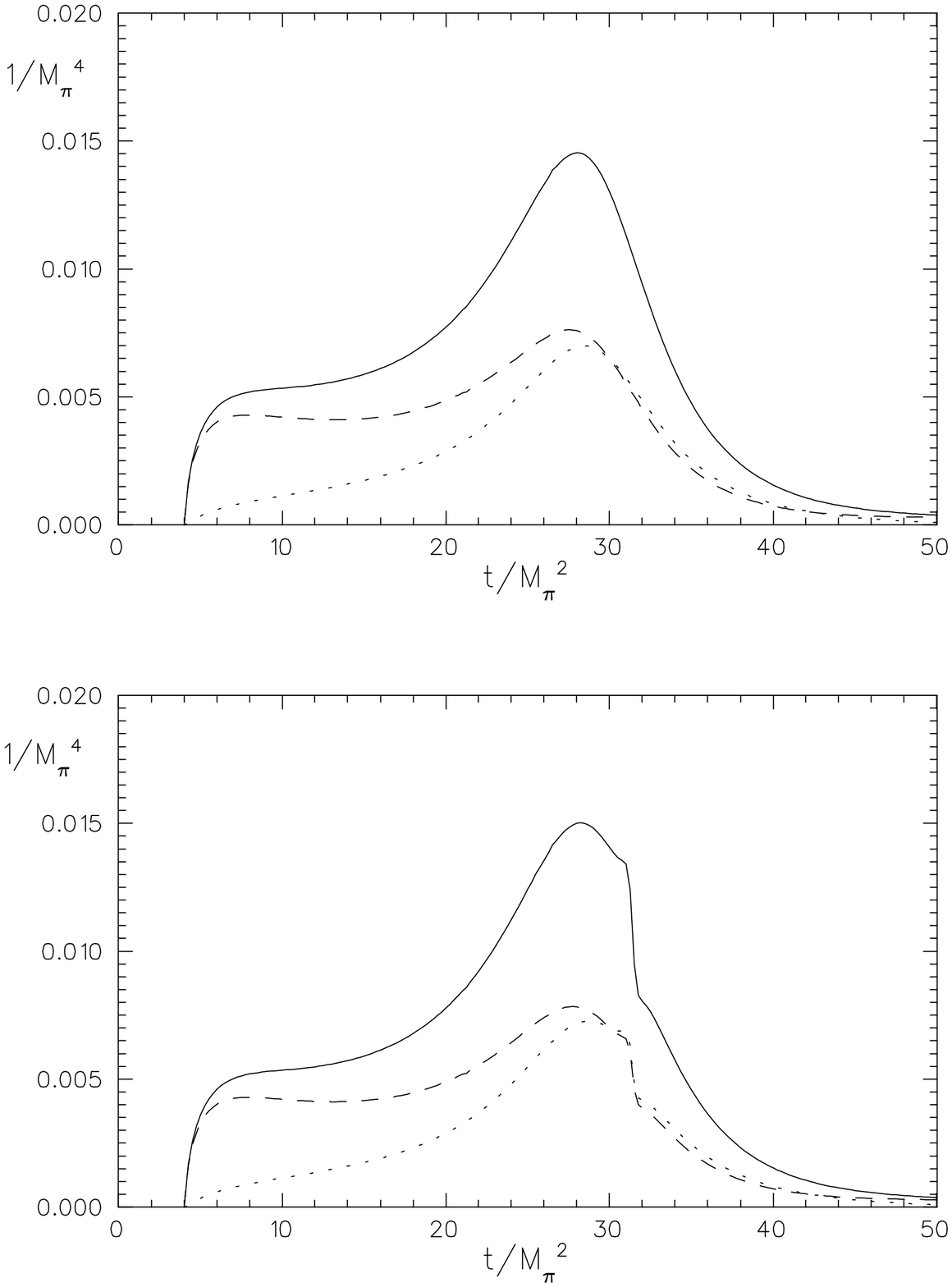}
}
\vskip 0.7cm

\centerline{\Large Figure 2}
\end{figure}

\newpage
$\;$\vspace{3cm}

\begin{figure}[bht]
\centerline{
\epsfxsize=5in
\epsfysize=7in
\epsffile{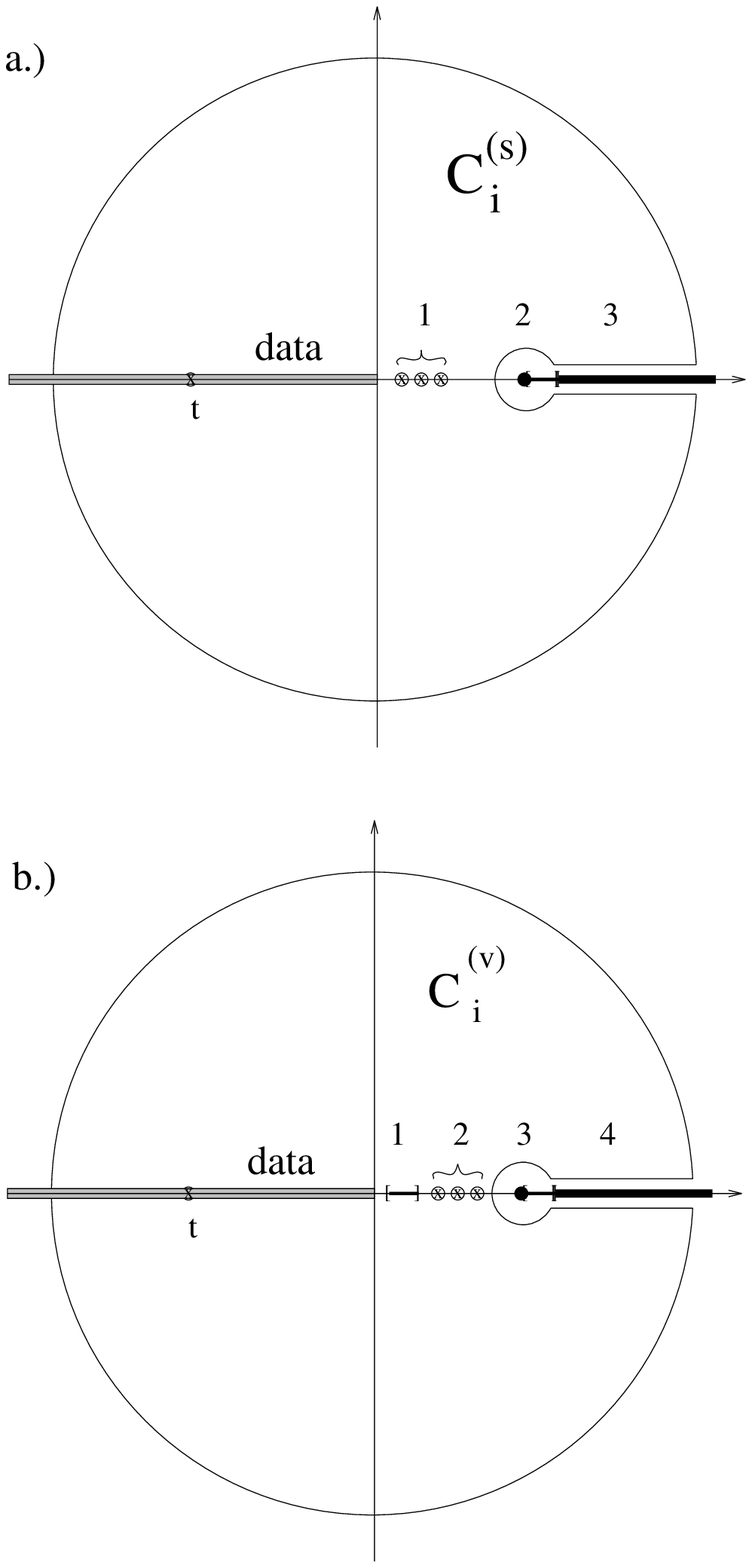}
}

\centerline{\Large Figure 3}
\end{figure}

\newpage
$\;$\vspace{3cm}

\begin{figure}[bht]
\centerline{
\epsfxsize=5.5in
\epsfysize=6.5in
\epsffile{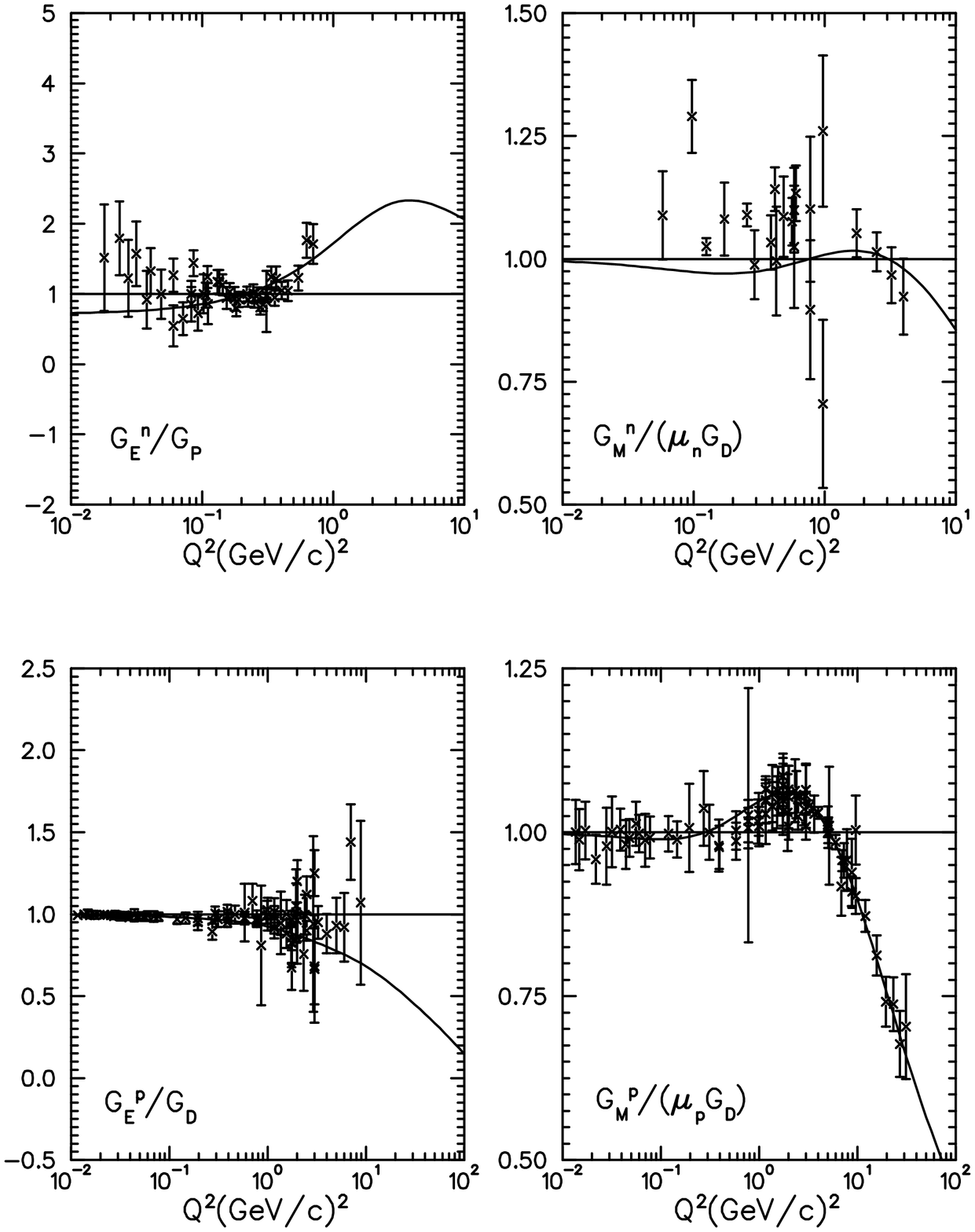}
}
\vskip 1.5cm

\centerline{\Large Figure 4}
\end{figure}

\newpage
$\;$\vspace{3cm}

\begin{figure}[b]
\centerline{
\epsfxsize=4.5in
\epsfysize=6in
\epsffile{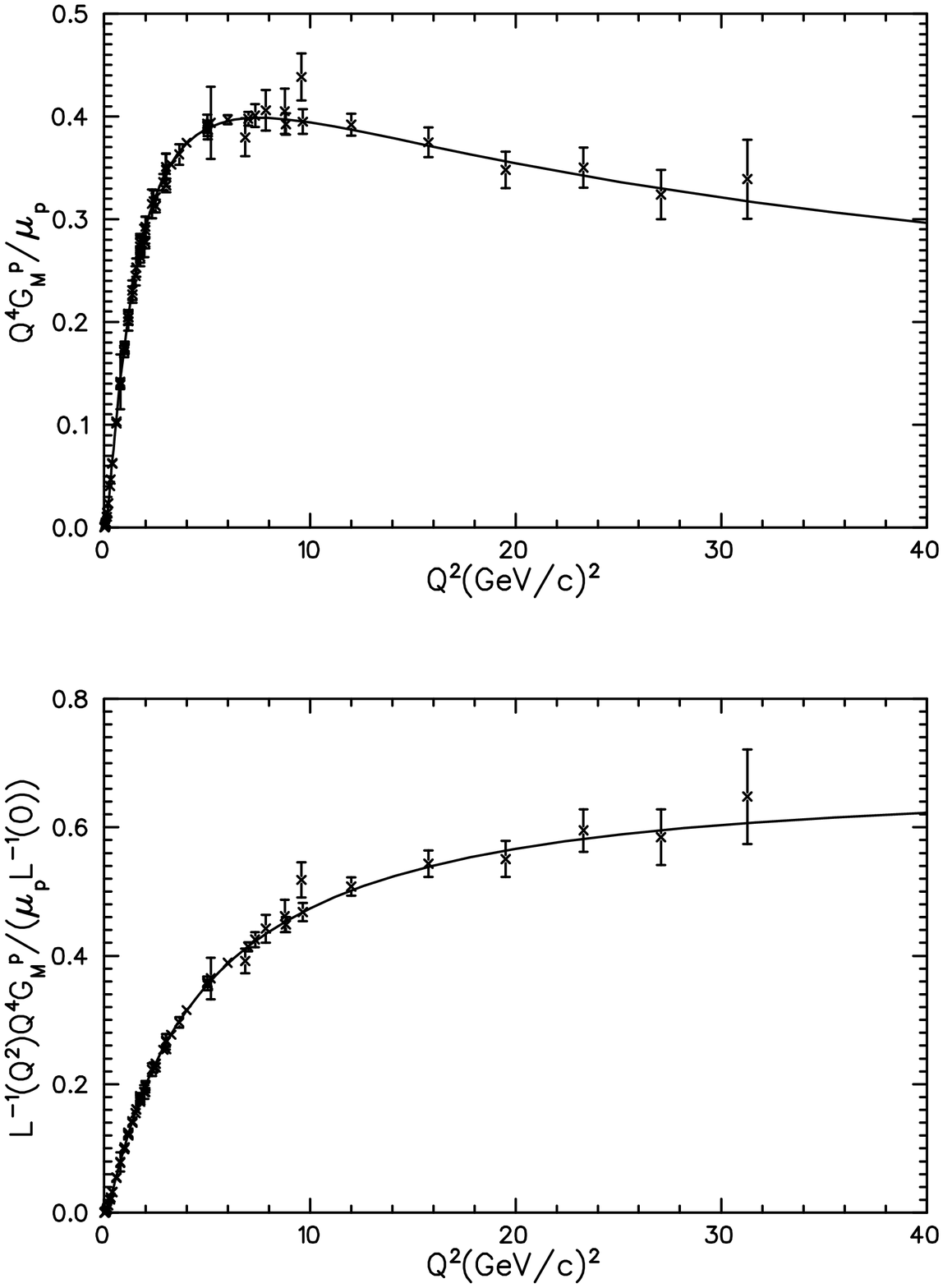}
}
\vskip 1.5cm

\centerline{\Large Figure 5}
\end{figure}

\newpage
$\;$\vspace{3cm}

\begin{figure}[bht]
\centerline{
\epsfxsize=5in
\epsfysize=6in
\epsffile{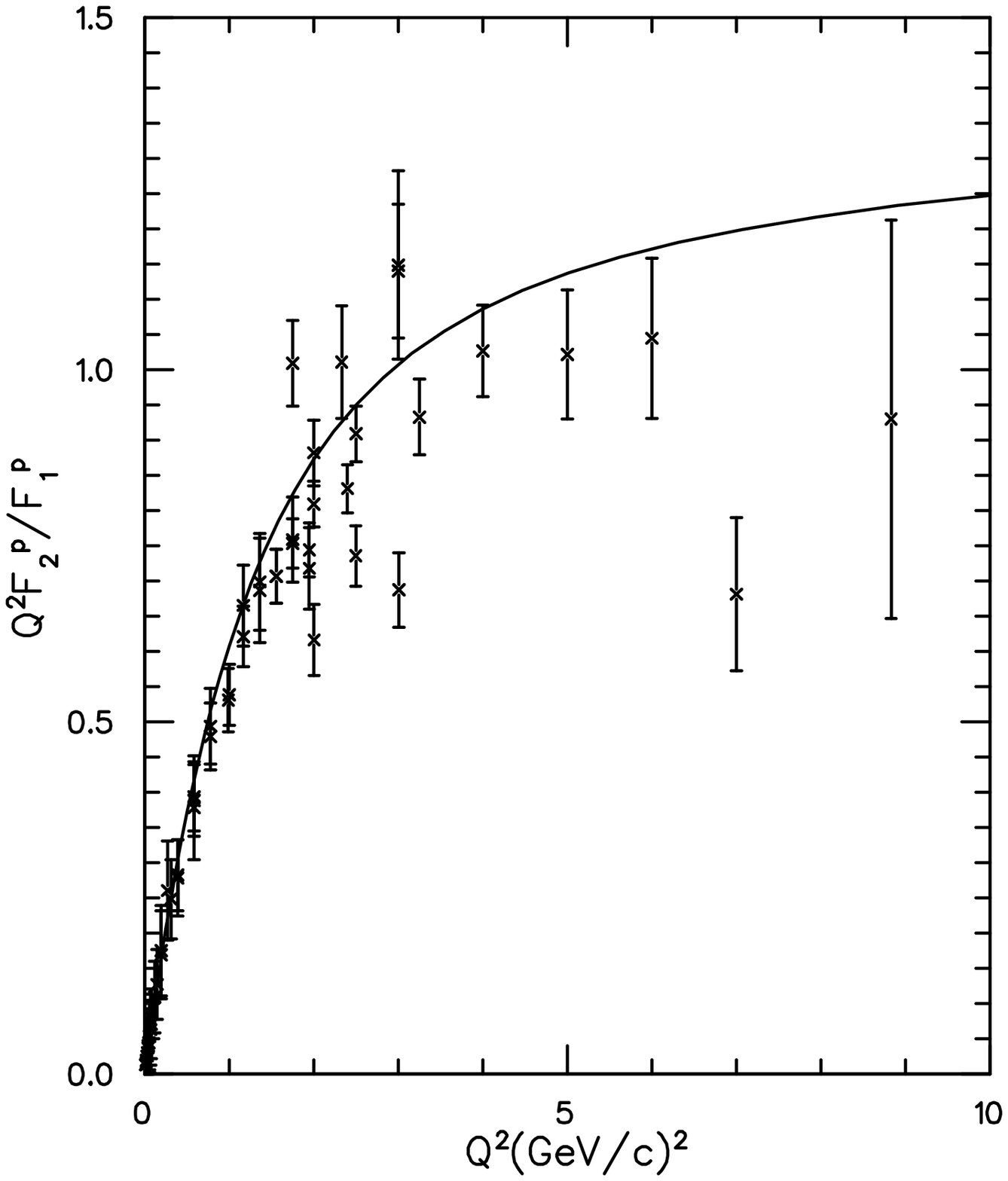}
}
\vskip 1.5cm

\centerline{\Large Figure 6}
\end{figure}
\newpage
$\;$\vspace{3cm}

\begin{figure}[bht]
\centerline{
\epsfxsize=5.5in
\epsfysize=6.5in
\epsffile{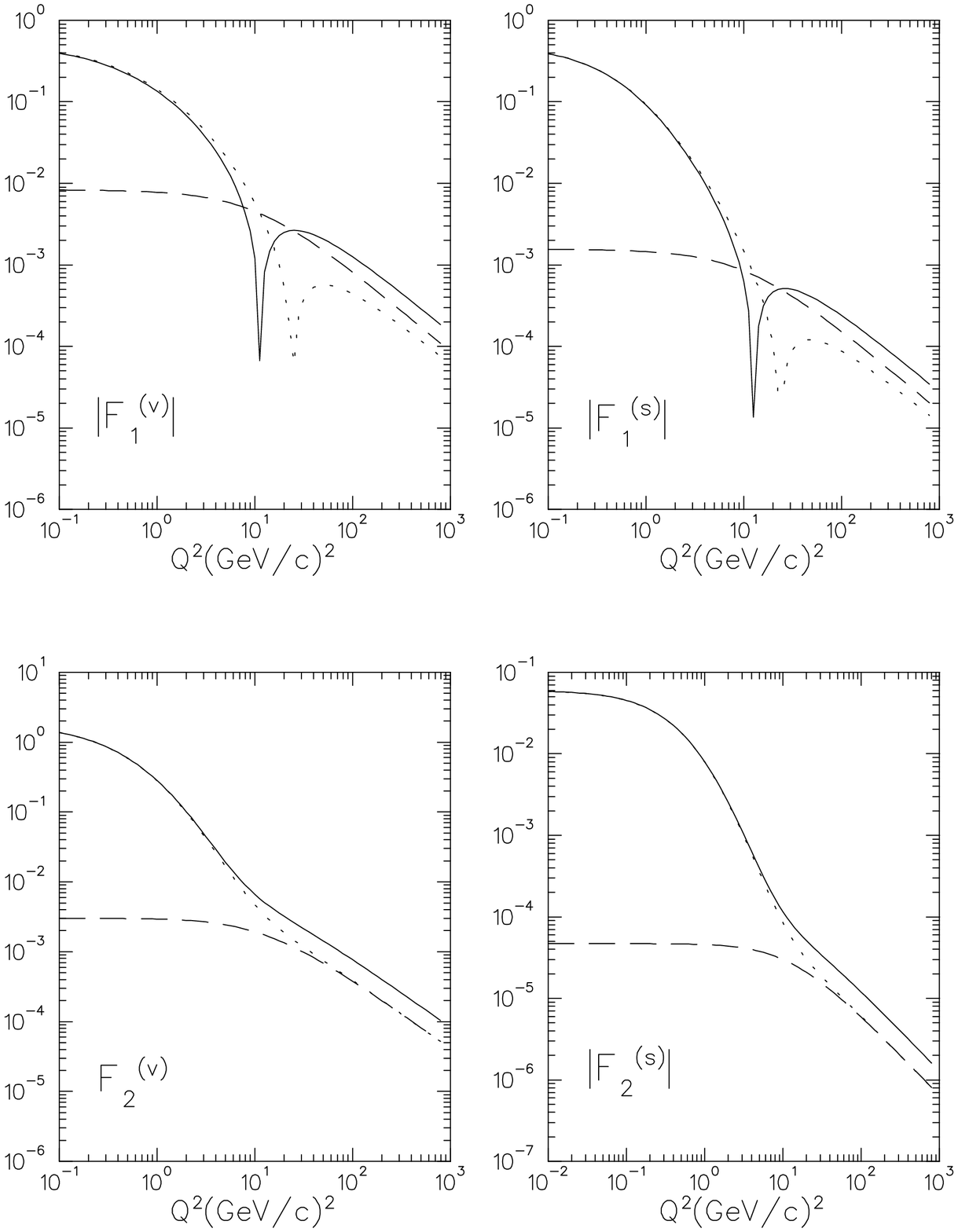}
}
\vskip 1.5cm

\centerline{\Large Figure 7}
\end{figure}

\end{document}